\documentclass[journal]{IEEEtran}

\pdfoutput=1

\usepackage{dblfloatfix}

\usepackage{times}
\usepackage{amsfonts}
\usepackage{amsmath}
\usepackage{verbatim}
\usepackage{epsfig}
\usepackage{enumerate}
\usepackage{xcolor,colortbl}
\usepackage{amssymb}
\usepackage{latexsym,color,amscd,textcomp,url}
\usepackage{rotating}
\usepackage{amsthm}

\newtheorem{theorem}{Theorem}

\usepackage{algorithm}
\usepackage{algpseudocode}
\usepackage{epsfig}
\usepackage{subfigure}
\usepackage{calc}
\usepackage{amsmath}
\usepackage{multirow}
\usepackage{adjustbox,lipsum}

\newcounter{countdefinitions}
\newcounter{counttheorems}
\begin{document}

\title{
Traffic models with adversarial vehicle behaviour
 }

\author{\IEEEauthorblockN{ Bogdan Groza} \\
\IEEEauthorblockA{Faculty of Automatics and Computers\\
Politehnica University of Timisoara, Romania\\
Email:  bogdan.groza@aut.upt.ro}
}

\maketitle

\begin{abstract}

We examine the impact of adversarial actions on vehicles in traffic.  Current advances in assisted/autonomous driving technologies are supposed to reduce the number of casualties, but this seems to be desired despite the recently proved insecurity of in-vehicle communication buses or components. 
Fortunately to some extent, while compromised cars have become a reality, the numerous attacks reported so far on in-vehicle electronics are exclusively concerned with impairments of a single target. In this work we put adversarial behavior under a more complex scenario where driving decisions deluded by corrupted electronics can affect more than one vehicle. Particularly, we focus our attention on chain collisions involving multiple vehicles that can be amplified by simple adversarial interventions, e.g., delaying taillights or falsifying speedometer readings. We provide metrics for assessing adversarial impact and consider safety margins against adversarial actions. 
Moreover, we discuss intelligent adversarial behaviour by which the creation of rogue platoons is possible and speed manipulations become stealthy to human drivers. 
We emphasize that our work does not try to show the mere fact that imprudent speeds and headways lead to chain-collisions, but points out that an adversary may favour such scenarios (eventually keeping his actions stealthy for human drivers) and further asks for quantifying the impact of adversarial activity or whether existing traffic regulations are prepared for such situations. 
\end{abstract}

\newcommand\true{\mathrm{True}}
\newcommand\false{\mathrm{False}}
\newcommand\ifff{\mathsf{~iff~}}

\newcommand\getx{\mathsf{get_x}}
\newcommand\gety{\mathsf{get_y}}
\newcommand\getdist{\mathsf{GetDist}}
\newcommand\getpos{\mathsf{GetPos}}

\newcommand\ilocked{\mathsf{HLocked}}
\newcommand\collision{\mathsf{Collides}}
\newcommand\iscrashed{\mathsf{Crashed}}
\newcommand\isbrakingorcrashed{\mathsf{BrakingOrCrashed}}
\newcommand\isbraking{\mathsf{Braking}}
\newcommand\isrunning{\mathsf{Running}}
\newcommand\getstate{\mathsf{State}}
\newcommand\paststate{\mathsf{OldState}}
\newcommand\stopsigs{\mathit{SSigs}}
\newcommand\stopvis{\mathsf{SVisible}}
\newcommand\vehicle{\mathit{car}}
\newcommand\intersection{\bigoplus}
\newcommand\squadron{\Sigma}
\newcommand\tmodel{\mathcal{M}}
\newcommand\lanes{\Lambda}
\newcommand\lane{\mathit{lane}}
\newcommand\real{\mathbb{R}}
\newcommand\startp{P_{\mathit{start}}}
\newcommand\stopp{P_{\mathit{stop}}}
\newcommand\direction{\chi}
\newcommand\angl{\phi}
\newcommand\dtime{\delta_{\mathit{react}}}
\newcommand\delay{\delta}
\newcommand\adv{\mathit{adv}}
\newcommand\pos{\mathit{pos}}
\newcommand\speed{\mathit{v}}
\newcommand\vlen{\mathit{vlen}}
\newcommand\visdist{\mathit{visd}}
\newcommand\vinit{v_{\mathit{init}}}

\newcommand\Max{\mathit{Max}}
\newcommand\stepf[1]{\theta(#1)}
\newcommand\stt{\mathit{state}}
\newcommand\crashed{\mathit{Crashed}}
\newcommand\tstamp{t_{\mathit{break}}}
\newcommand\running{\mathit{Running}}
\newcommand\braking{\mathit{Braking}}
\newcommand\dist[2]{|#1 - #2|}

\newcommand\dbreak{d}
\newcommand\ccount{\lambda}
\newcommand\fspeed{\vartheta}
\newcommand\freact{\epsilon}
\newcommand\nttr{\mathrm{irc}}
\newcommand\ratio{\rho}

\newcommand\lsec{\ell}

\section{Introduction and motivation}

Due to the high relevance for modern society, preventing and modelling traffic collisions has been a constant research preoccupation in the past few years.
There is a significant number of publications on this topic and many of them particularly address chain-reaction crashes (car pile-ups). For example,
Android-based prototype implementations for collision avoidance are discussed in \cite{Chen16}. Preventing pile-up crashes in platoons where only part of the vehicles are equipped with advanced warning capabilities is accounted in \cite{Chakravarthy09}.
Stochastic models for chain collisions are studied in \cite{Garcia12} and \cite{Garcia13}. More accurate models for the estimation of crash probabilities based on vehicle trajectory for autonomous driving are discussed in \cite{Althoff09}. Platoons with various penetration rates of inter-vehicle communication units are taken into account in \cite{Ttian16}.

Since the first comprehensive security analysis of modern vehicles in \cite{Koscher10} and \cite{Checkoway11}, dozens of attacks on in-vehicle electronics are reported each year proving a high degree of insecurity.  Consequently, adversarial vehicle behaviour is as realistic as possible. Dozens of works focused on assuring the security of in-vehicle buses, e.g., \cite{Groza13}, \cite{Lin15}, \cite{Woo15}, \cite{Woo16}, did not receive enough echo from the industry as none of the vehicles on road today attains the necessary security level. This makes vehicles trivial targets for determined adversaries.

Still, there is little attention focused on adversarial vehicle behaviour, i.e., vehicles that are compromised by malicious adversaries and misbehave while in traffic, deluding the driver and other traffic actors, potentially leading to serious traffic incidents that involve multiple vehicles, e.g., chain collisions. 
The traditional adversarial setup for in-vehicle communication assumes an adversary that tampers with data on insecure buses resulting in malfunction of the vehicle, e.g., stopping the engine, killing the brakes, etc. 
Such attacks are easy to attain as long as existing in-vehicle buses, e.g., the Controller Area Network (CAN), FlexRay or BroadR-Reach (an Ethernet based technology), are lacking security mechanisms. All the attacks reported so far were performed in isolated environments and rarely on road, e.g., the Jeep hack incident\footnote{https://www.wired.com/2015/07/hackers-remotely-kill-jeep-highway/}. Moreover, damages are generally restricted to a single target vehicle. 
In contrast, the view expressed by our work accounts for the possibility of more than a single target vehicle and opens road for more complex scenarios. 
 
\emph{Structure of our work.}
 For clarity,  the main ideas of our work can be summarized as follows:
 
 \renewcommand\labelitemi{$\bullet$}
 
\begin{itemize}
\item  we emphasize on a view that stems from the \emph{driver-vehicle-environment} system and sets stage for  \emph{adversarial vehicle behaviour} by which vehicles may misbehave, e.g., delaying taillights, displaying false speedometer readings, etc., (Section I), 

\item  we discuss models for chain collisions in the presence of adversarial vehicle behaviour and provide two metrics for assesing the impact: \emph{the infinite collision bound} and the \emph{instant-reaction-collision speed gain} (Section II),

\item we discuss \emph{safety margins against adversarial behaviour} in an attempt to determine how existing safety rules (such as the 2-second rule) translate in the presence of adversarial vehicle behaviour (Section II),

\item we provide simulations as overlays on existing maps in order to gain a more realistic feeling and some experimental data (Section III),

\item we discuss two forms of intelligent adversarial behaviour:  \emph{adversarial platoon formation} by which an adversary manages to coagulate multiple vehicles and \emph{stealthy speed manipulations} that will allow an adversary to progressively modify the speed of the car without being noticeable for human drivers (Section IV).

\end{itemize}

\begin{figure*}[thb!]
\begin{center}
\includegraphics[width=14cm]{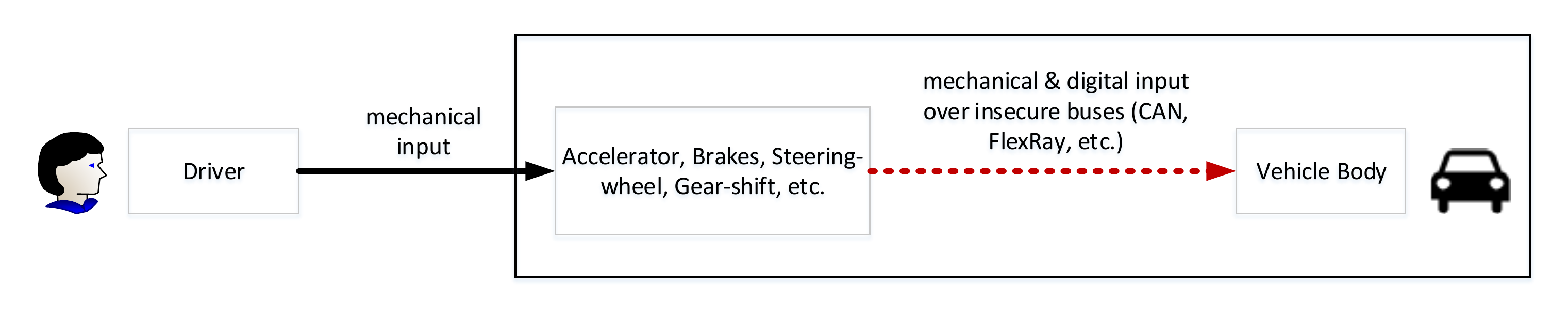} 
\end{center}
\caption{The naive (open-loop) view of the driver-vehicle system: the driver exerts mechanical input over the vehicle body, mediated by potentially insecure communication on vehicle buses}
\label{fig:env_classic}
\end{figure*}

\subsection{An extended view: the driver-vehicle-environment system}

The classical view over automotive security assumes the existence of a corrupted in-vehicle network (controlled by an adversary) that mediates the interaction between the driver and the vehicle. This may suggest the naive open-loop image over the driver-vehicle system that is suggested in Figure \ref{fig:env_classic}.

The view that we advocate extends the simple open-loop system from Figure \ref{fig:env_classic} to a more complex closed-loop environment which is closer to the real-world model. In this setup, the driver exerts mechanical input over the vehicle body which in turn impacts the environment. Further,  the driver continuously receives acoustic and visual inputs from both the vehicle and the environment. 
At least part of the driver actions are mediated by the insecure communication from the in-vehicle buses. But the adversarial nature of the vehicle also extends to the environment. This is justified as other traffic participants may be similarly corrupted vehicles that behave dishonestly by delaying taillights, disabling side-lamps, etc.  
The closed-loop view of the \emph{driver-vehicle-environment} triplet is suggested in Figure \ref{fig:env_enhanced}. While our work generally refers to human drivers, we make it explicit that in Figure \ref{fig:env_enhanced} the role of the human driver (1) can be played by some autonomous driving module (1'). Mutatis mutandis, our results can be easily re-interpreted in the context of autonomous driving since electronic devices may take similar decisions by interpreting visual and acoustic signals from the environment.

The driver-vehicle-environment triplet forms a complex system where adversarial behaviour on various components can have serious consequences over multiple participants rather than restricted effects on a single vehicle/driver. The relevance of this broader image stems from the impact on other road participants and opens the possibility for chain reactions that put the problem at a larger scale involving hundreds of cars rather than a single participant.

\begin{figure*}[thb!]
\begin{center}
\includegraphics[width=15cm]{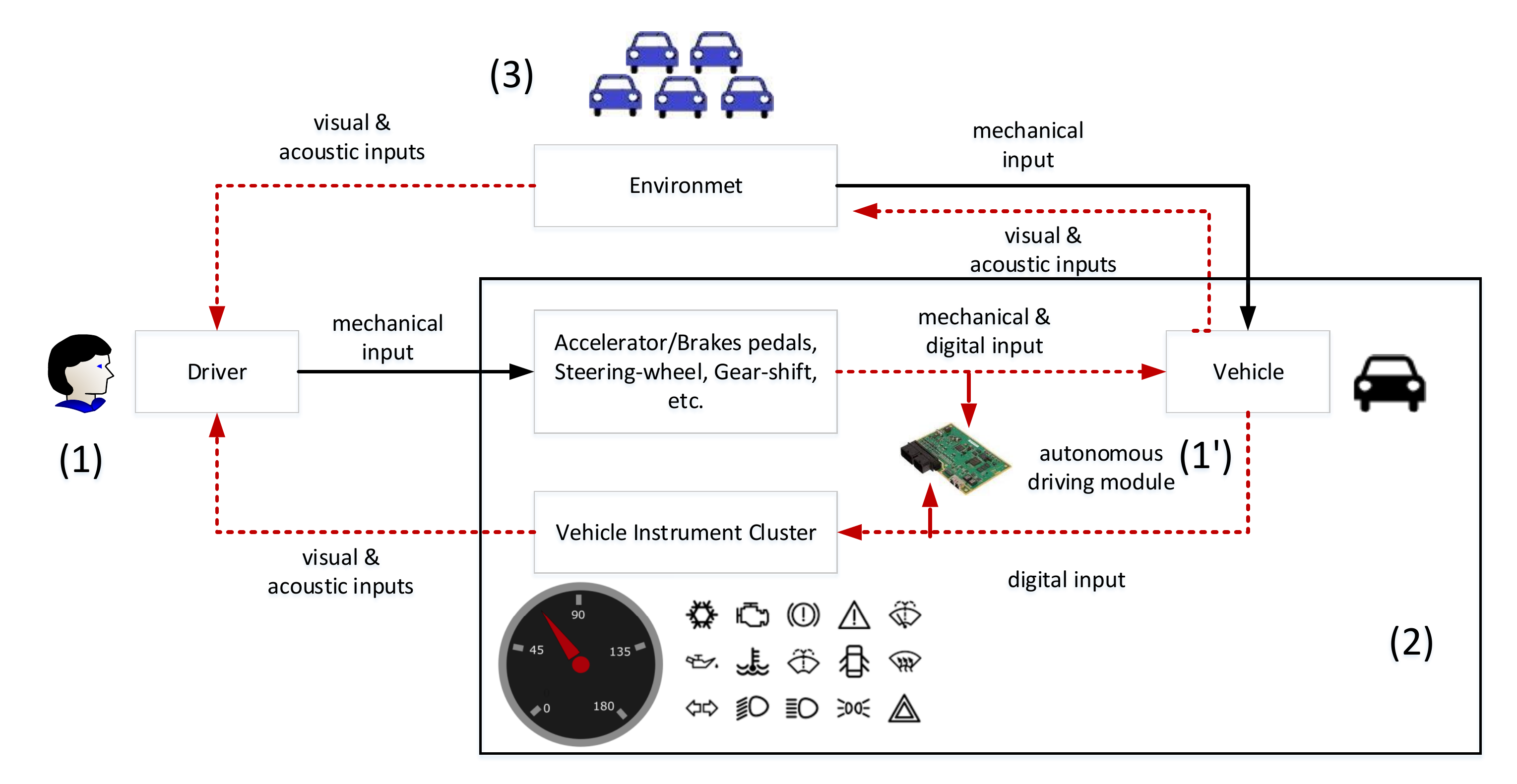} 
\end{center}
\caption{The enhanced (closed-loop) view of the driver-vehicle-environment system: the driver (1) receiving inputs exerts mechanical actions over the vehicle body (2) which in turn impacts the environment (3), all these mediated  by potentially insecure communication}
\label{fig:env_enhanced}
\end{figure*}

There are many factors influencing traffic safety, including driver's behaviour, the environment or vehicle condition, etc. Some of these are in immediate reach for manipulation by an adversary. In a comprehensive study on speed and safety \cite{Feng01} published more than a decade ago, five factors are taken into account, all of them are relevant to the context and models addressed by our work:

\begin{enumerate}
\item \emph{headway} - the distance between cars at a given speed which is the key factor in chain-collisions,

\item \emph{vehicle speed and speed limitations} - which are likely the main factor in increasing or reducing traffic casualties,

\item \emph{environment} - which not only dictates the safety speed and headway but can also become adversarial in the context addressed here (e.g., braking or turning without signalling),

\item \emph{distractions} - besides regular phones or smart-phones, modern cars have complex infotainment units and media streaming services that can distract the driver even more.

\end{enumerate}

To bring more context to the problem we give a brief account of driver behavior and existing regulations.

\subsection{Driver behaviour and regulations}

Assessing the real-world impact requires a crisper image over the driver behaviour and perception.
To get a more realistic view on the context in which hazardous situations take place, it is useful to caps on the following two recommendations that serve as heuristics for most drivers:

\begin{enumerate}
\item \emph{Drive only so fast that the vehicle is under control}.
There is general consensus that one should drive a vehicle only so fast that the vehicle is still under control - we will call this \emph{the safety rule}. But drivers are not always prudent and accidents due to speeding are still numerous, which proves that this recommendation is either disregarded or incorrectly used. Moreover, it turns out that drivers are often wrong in assessing the speed at which the vehicle is controllable. This is proved both by studies which show driver inaccuracies in predicting the speed but also by statistics which commonly points out that speed limitations do greatly reduce the number of casualties (which implies that drivers fail in estimating the safety speed). According to some of the results summarized in \cite{Stuster98}, decreasing the speed limit from 110 km/h to 90 km/h in Sweden lead to 21\% decrease in fatal crashes, while in Germany decreasing from 60 km/h to 50 km/h lead to a decline in crashes by 20\%, etc.
Since prior to such speed limitations, drivers did have in mind the safety rule, 
it means that drivers are not that good in establishing the safety speed and take the legal maximum for granted. 

\item  \emph{The 2 seconds rule or keep apart 2 chevrons}. The recommendation that the driver should stay 2 seconds away from the vehicle in front seems to be generally accepted in most European countries as well as in the US \cite{CEDR09}. The first problem with this rule is that 2 seconds cannot guarantee a safe stopping distance (see Table \ref{tab:distance}) and  can generally  cover only the driver reaction time which is at around 1.5 second. Another problem is that drivers perception of distance to objects may not be very accurate. 
Commonly, the EU or US highways (the most common place for car piles) require drivers to keep 2 chevrons between cars. Chevrons are graphically depicted on the road and announced by sideways markings as suggested in Figure \ref{fig:2chevrons}.  This is known, and proved by scientific evidence, to reduce the number of accidents. In \cite{Greibe10} chevrons spaced by 36m (i.e., the 2-seconds bound) are reported to reduce the speed by 1-3km/h and the number of vehicles with less than 1s headway. 
\end{enumerate}

\begin{figure}[thb!]
\begin{center}
\includegraphics[width=5cm]{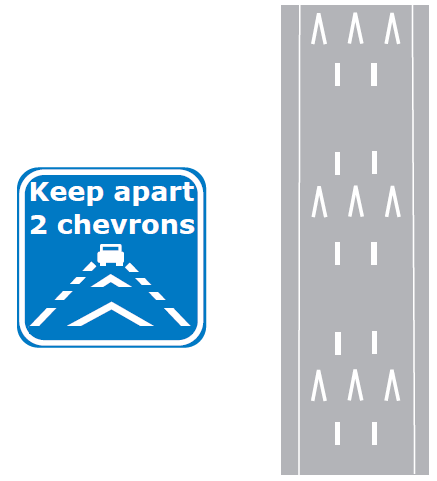} %, angle =90
\end{center}
\caption{Suggestive depiction of the chevrons highway marking}
\label{fig:2chevrons}
\end{figure}

As proved by practical incidents, these heuristics are still far from keeping accidents away and under adversarial vehicle behavior the situation is even worse as we discuss in the following sections. 

\subsection{Fixed parameters in our models}

To serve as ground for a quantitative approach, the following generally accepted numerical values are considered by us as well:

\begin{enumerate}
\item  \emph{Driver reaction time} is generally considered as 1.5 seconds. Experienced drivers commonly react at under 1 second while for elders 2 seconds are more realistic. 
Standard reaction time can be greatly impaired in adversarial conditions if the taillights are disabled since driver decisions rely on the perceived distance to the car in front. 

\item  \emph{Speed regulations} may tempt the driver to take the car to the limit rather than adjust its speed according to the 2 seconds rule or the safety rule. Commonly accepted maximum speed limitations include 50 km/h in cities, 100 km/h outside urban areas and 130 km/h on highways.

\item  \emph{Kinetic friction coefficient} is usually taken at 0.7 for accident reconstructions. This can of course vary for icy or wet surfaces but it is beyond the scope of our presentation to consider such variations.
\end{enumerate}

In Table \ref{tab:distance} we include some values for the braking distance under various conditions. It can be easily seen that the sum between the distance caused by driver's reaction time and the braking distance quickly exceeds the distance travelled by the car in 2 seconds. The 2-seconds rule has its limitations and mostly works if the obstacle in front is also a braking vehicle but cannot compensate in case of an immediate obstacle. Another problem of the 2-chevrons rule in the context of adversarial vehicle behavior is that chevrons are spaced assuming a speed of 130 km/h, but if vehicle's speed/speedometer is manipulated the space between 2 chevrons (72m) no longer corresponds to 2-seconds safety distance.

\begin{table}[h]
\begin{center}
\caption{Distance due to reaction time, braking and the 2-seconds rule}
\label{tab:distance}
  \begin{tabular}{ c  c c c c c }
    \hline
 speed (km/h) & 20 & 30 & 50 & 90 & 130 \\ \hline \hline
    \emph{distance at 1.5s reaction time (m)} & 8.3 & 12.5 & 20.8 & 37.5 &  54.1 \\ 
 \emph{braking distance (m)} & 2.2 & 5 & 14 & 45.5 & 95 \\ 
  \emph{distance in 2s (m)} & 11.1 & 16.6 & 27.7 & 50 & 72.2 \\
    \hline
  \end{tabular}
 \end{center}  
\end{table}

\section{Models and adversarial behavior for multiple vehicle collisions}

We begin by presenting a simple model for multiple vehicle collisions then we add adversarial actions and discuss impact on the model. Table \ref{tab:notations} provides a summary for the notations that we use in this section.

\subsection{Model for single lane multiple vehicle collision}

In Figure \ref{fig:veh_lane} we present vehicles on a lane, the length of a vehicle is $l$ and the headway (space between vehicles) is $b$. We proceed by modelling multiple collisions on a single lane followed by an intersection and then we add adversarial behaviour to these models. 

\begin{figure}
\begin{center}
\includegraphics[width=9cm]{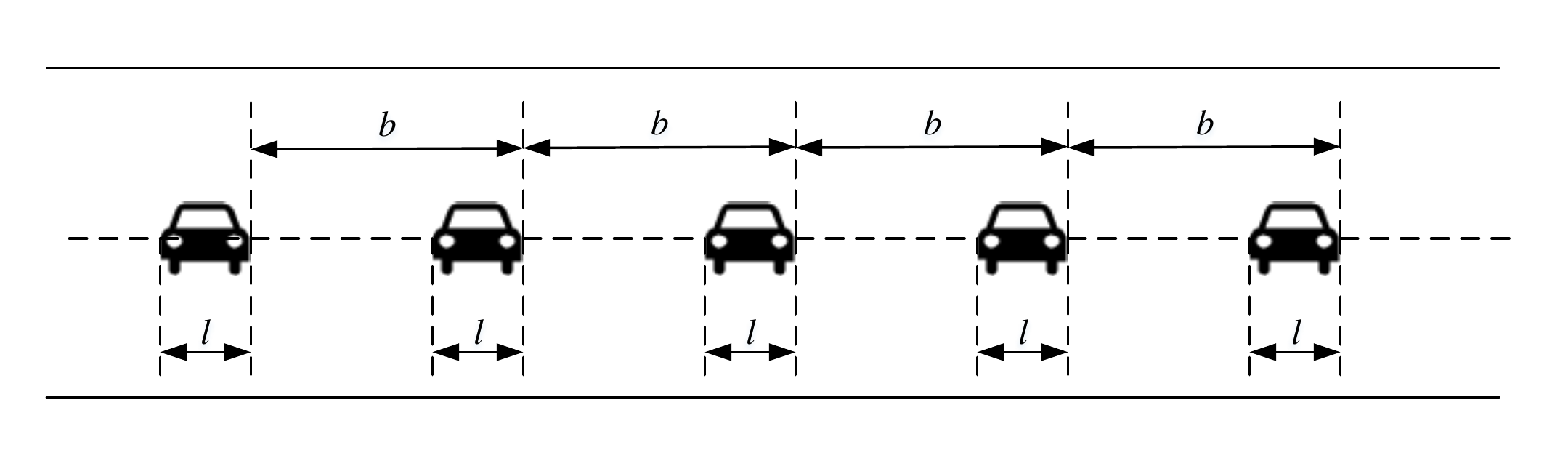} %, angle =90
\end{center}
\caption{Vehicles with length $l$ at headway $b$}
\label{fig:veh_lane}
\end{figure}

The dynamics of vehicle braking are well understood. We simply compute the braking distance by equating the kinetic energy with the work done by braking, i.e.,

\begin{equation}
\frac{1}{2} m v^2=\mu  m  g  d
\end{equation}

Here $m$ is vehicle's mass, $v$ its speed, $\mu$ is the friction coefficient, $g$ is the acceleration due to gravity and $d$ is the braking distance. The braking distance directly follows as:

\begin{equation}
d = \frac{v^2}{2  \mu  g}
\end{equation}

\begin{table}[t!]
\begin{center}
\caption{Summary of notations}
\label{tab:notations}
  \begin{tabular}{ c | l}
  \hline \hline 
  $g$ & acceleration due to gravity \\
  $m$ & vehicle mass \\
  $\mu$ & friction coefficient \\
  $\speed$ & vehicle speed \\
  $b$ & headway \\
  $\dbreak$ & braking distance \\
  $l$ & vehicle length \\
  $\dtime$ & driver reaction time \\
  $\fspeed$  & speed increase due to adversarial manipulation \\
  $\freact$  & decrease in reaction time due to adversarial intervention \\
  $\speed_{\adv}$  & total reaction time after adversarial intervention \\
  $\delay_{\adv}$  & vehicle speed after adversarial intervention \\
  $\ccount$ & number of crashed vehicles \\
  $\ell$ & safety headway in front of adversarial manipulation \\
    \hline
  \end{tabular}
 \end{center}  
\end{table}

These equations alone are sufficient to assess the severity of an impact in case of vehicles at headway $b$ as depicted in Figure \ref{fig:veh_lane}.  The condition for the $\ccount$-th vehicle to collide follows by requiring the distance to the vehicle in front, i.e., $\ccount  b$, to be smaller than the braking distance plus the reaction time, i.e., $\dbreak + \speed   \ccount   \dtime$. This leads to:

\begin{equation}
\ccount  b <  \frac{v^2}{2  \mu  g} + \ccount \speed \dtime
\end{equation}

We assume that $\speed   \dtime < b$, since otherwise the headway is too short, i.e., it takes too long for the driver to hit the brakes, and immediate collision occurs. Then we easily obtain an upper bound for $\ccount$:

\begin{equation}
\ccount  <  \dfrac{ \dfrac{v^2}{2   \mu  g}}{b-\speed  \dtime}
\label{eq:count}
\end{equation}

When $b = \speed   \dtime$ the number of collisions tends to infinity. An infinite number of collisions should not come as a surprise since chain collisions involving more than one hundred cars occurred on several occasions in the real-world (e.g., a 259 car pileup stretching over 30 km happened on the German autobahn A2 in 2009\footnote{https://www.thelocal.de/20090720/20701}). These real-world numbers can get worse as in theory the number of collisions tends to infinity when $b$ approaches $\speed   \dtime$.

\subsection{Adversary capabilities and impact of adversarial behaviour}

There are a number of actions that can be taken by an adversary, but in our model we do focus on two actions that may not be even noticeable to the driver:
\begin{itemize}
\item \emph{Falsifying speedometer readings} which will likely misled the driver to run at a distinct speed. If the speedometer presents false readings indicating a lower speed, the driver will go faster, rather than assuming that the speedometer is wrong. External readings from an uncompromised device, e.g., some GPS software from the mobile phone can alert the driver on a potential malfunction, but such situations are out of scope for this work (it is unlikely that all drivers will rely on external measurements and even these can be compromised). We modify regular vehicular speed by $\fspeed$:

\begin{equation}
\speed_{\adv} \leftarrow \speed + \fspeed
\end{equation}

\item \emph{Delaying reaction time} directly translates in adding an adversarial delay to braking or to vehicle taillights. If brakes are controller by electrical means, i.e., brake-by-wire systems which are tentative replacement for mechanical systems in the near future, such delays can be forced by simply delaying messages on the bus. However, even for mechanical systems the adversary can indirectly delay the reaction time of the driver from behind by delaying the taillights. Taillights were previously considered in modelling multiple vehicle collisions \cite{Nagatani15} and clearly they are a common source of accidents. Several studies show that faster LED stop lamps are more effective than light bulbs in reducing the number of collision (but these seem controversial \cite{Greenwell13}). A fundamental work in the visual control of braking \cite{Lee76} points out that if the lead vehicle is without braking lights, the reaction time can be longer than 2 seconds. This result is relevant as it clearly renders the 2-second rule ineffective when taillights are manipulated by adversaries. We consider that adversarial actions in delaying or disabling the taillights result in a delay $\freact$ added to driver reaction time:

\begin{equation}
 \delay_{\adv} \leftarrow \dtime + \freact
\end{equation}

\end{itemize}

\emph{Impact.} We consider useful to introduce the following two metrics for adversarial capabilities:

\begin{itemize}

\item The \emph{$\infty$-collision bound} is defined by the set of pairs $(\fspeed, \freact)$ for which collision of an infinite number of vehicles occurs. The dependence between $\fspeed$ and $\freact$ can be easily computed from the $\infty$-collision condition, i.e., $b=\speed_{\adv}  \delay_{\adv}$, as:

\begin{equation}
\freact(\fspeed) = \dfrac{b}{\speed + \fspeed} - \dtime
\end{equation}

\item  The \emph{instant-reaction-collision speed gain} $\fspeed_{\nttr}$ is the speed induced by an adversary for which the driver cannot stop the vehicle even if it instantly reacts to front-vehicle braking. Assuming no adversarial delays and a 2-second headway we have:

\begin{equation}
 \fspeed_{\nttr} = 2 \sqrt{\strut \speed \mu g} -\speed 
\end{equation}

This follows from the fact that the \emph{instant-reaction-collision speed gain} requires the braking distance to be equal to the headway:

\begin{equation}
b = \dfrac{\speed_{\adv}^2}{2   \mu   g}
\end{equation}

In case of a 2-second rule headway as $b = 2   \speed$ and $\speed_{\adv} \leftarrow \speed + \fspeed$ it follows:

\begin{equation}
 2   \speed = \dfrac{\speed^2 +2\speed\fspeed+ \fspeed^2}{2   \mu   g} \Rightarrow \fspeed_{\nttr} = 2 \sqrt{\strut \speed \mu g} -\speed 
\end{equation}

\end{itemize}

To clarify this by a practical example consider the regular highway speed $\speed=130km/h$. Following the two seconds rule (which was already proved not to be very efficient for this case) we have a headway $b=72 m$. We discuss impact on graphical representations. 

On the left side of Figure \ref{fig:react_adv} we depict the impact of speed modifications on the number of collisions. At $30 km/h$ there are already more than 25 vehicle that collide. On the right side of Figure  \ref{fig:react_adv} we depict the impact of modifications in the reaction time. A small delay of $400 ms$ is sufficient to lead to more than 25 vehicle collisions. In both situations the number of vehicles that collide grows drastically.

The left side of Figure \ref{fig:inf_coll_v_incontrol} depicts the infinite collision bound in relation to falsified speed and delayed reaction. Then on the right side of Figure \ref{fig:inf_coll_v_incontrol} we show the \emph{instant-reaction-collision speed gain} in relation with vehicle's speed at a 2 second headway.

\begin{figure}
\begin{minipage}{4.25cm}
\includegraphics[width=4.25cm]{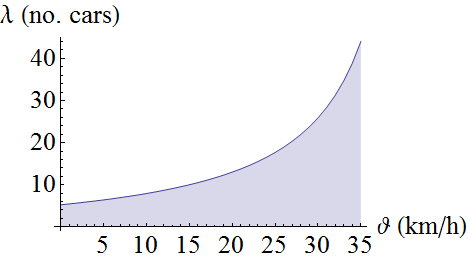} %, angle =90
\end{minipage}
\begin{minipage}{4.25cm}
\includegraphics[width=4.25cm]{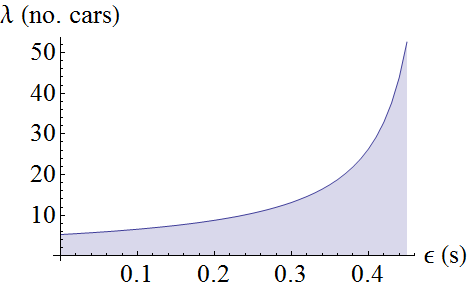} %, angle =90
\end{minipage}
\caption{Increase in the number of collisions with speed modification (left) and with delayed reaction time (right) (at $\speed=130km/h$ and $b=72 m$ based on the 2 seconds rule)}
\label{fig:react_adv}

\begin{minipage}{4.25cm}
\includegraphics[width=4.25cm]{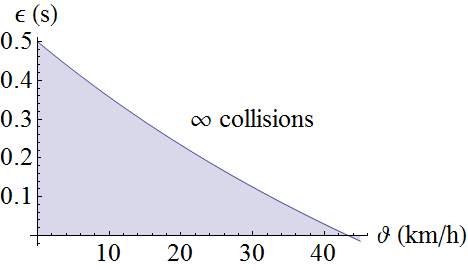} %, angle =90
\end{minipage}
\begin{minipage}{4.25cm}
\includegraphics[width=4.25cm]{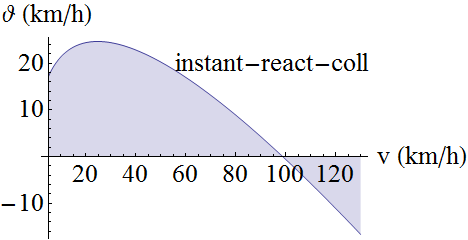} %, angle =90
\end{minipage}
\caption{The $\infty$-collisions bound (left) at $\speed=130km/h$, $b=72 m$ based on the 2 seconds rule and speed increase due to adversarial action to render braking out of control (right)}
\label{fig:inf_coll_v_incontrol}
\end{figure}

Figure  \ref{fig:speed_react_adv} combines modifications in speed and reaction time in a 3D plot. A number of more than 100 collisions is quickly reached. Then in Figure \ref{fig:inf_coll_3d} we depict the $\infty$-collisions bound. In theory when $b = \speed   \dtime$ an infinite number of collisions occur. This means that at speed $\speed=130km/h$ with the two second rule headway $b=72 m$ a reaction time of $1.999$ leads to an infinite number of collisions. Similarly, a speed of $172 km/h$ leads to an infinite number of collisions.

\begin{figure}
\begin{center}
\includegraphics[width=8.5cm]{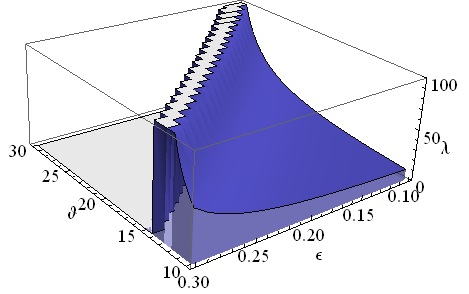} %, angle =90
\end{center}
\caption{Increase in the number of collisions with speed modification and delayed reaction time ($\speed=130km/h$ and $b=72 m$ based on the 2 seconds rule)}
\label{fig:speed_react_adv}
\begin{center}
\hspace*{-0.4cm}\includegraphics[width=10.4cm]{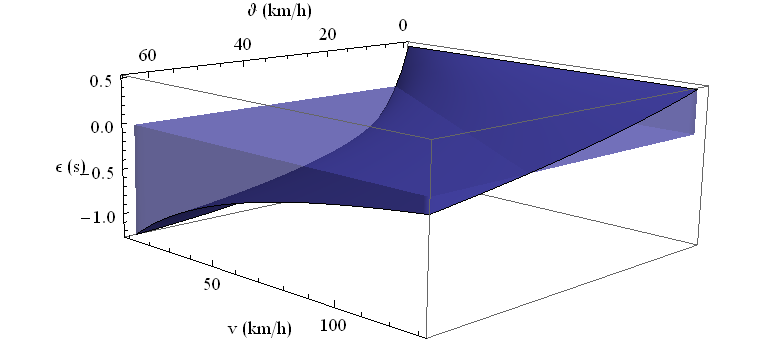} %, angle =90
\end{center}
\caption{The infinite collision bound for $\speed \in [0, 130]$ (km/h)}
\label{fig:inf_coll_3d}
\end{figure}

\subsection{Safety margins against adversarial behavior}

We now try to determine new safety rules following the potential impact of adversarial behaviour. Briefly, assuming no adversarial delay for taillights, we determine that in case of speed manipulations of at most $50\%$ (i.e., $\fspeed/\speed=0.5$) the \emph{2-second} rule translates to \emph{2-seconds plus 4\% of vehicle speed (in km/h)}. To state it otherwise, this means \emph{2 seconds plus 1 second for each 25 km/h} for a safe braking distance between vehicles. We explain this result in what follows.

Assume that an $\lsec$-seconds headway, i.e., $b=\lsec   \speed $, is safe, then:

\begin{multline}
\lsec  \speed > \dtime  \speed_{\adv}+ \dfrac{\speed_{\adv}^2}{2  \mu  g} \\ 
\Rightarrow \lsec  \speed > \dtime (\speed+\fspeed) + \dfrac{(\speed+\fspeed)^2}{2  \mu  g} \\
\Rightarrow \lsec > \dtime (1+\dfrac{\fspeed}{\speed}) + \dfrac{(\speed+\fspeed)^2}{2 \speed  \mu  g} 
\end{multline}

As already mentioned, for accident reconstruction $\dtime = 1.5$, $\mu=0.7$ are the norm. Since $g=9.8$ it follows:

\begin{equation}
\lsec > 1.5 (1+\dfrac{\fspeed}{\speed}) + 0.07 \dfrac{(\speed+\fspeed)^2}{\speed } 
\end{equation}

In Figure \ref{fig:speed_mspeed_adv} we graphically depict modifications of the safety distance $\lsec$ (expressed in seconds) in relation to vehicle reported speed $\speed$ and actual modifications by the adversary $\fspeed$. Generally, adversarial manipulation increases the safety margin from 2--3 seconds up to 4--6 seconds.

Now we consider the adversarial speed modification $\fspeed$ as some ratio $\ratio$ of the vehicle reported speed $\speed$, i.e., $\fspeed = \ratio \speed$. It follows that:

\begin{equation}
\lsec > 1.5 +1.5\ratio + 0.07 (\ratio+1)^2 \speed 
\end{equation}

For a more convenient interpretation, since in the previous relation speed was expressed in $m/s$, to convert to the speedometer scale in $km/h$ we multiply by $\dfrac{1000}{3600}=0.27$ which leads to:

\begin{equation}
\lsec > 1.5 +1.5\ratio + 0.019 (\ratio+1)^2 \speed '
\end{equation}

At a ratio $\ratio$ of at most $50\%$ we have an approximate minimum safety distance of $\lsec \approx 2 + 0.04\speed '$ and hence \emph{2 seconds plus 1 second for each 25 km/h}.

We now consider the impact of adversarial manipulation of reaction time. Relation (11) now translates to:

\begin{equation}
\lsec > (1.5+\freact) (1+\dfrac{\fspeed}{\speed}) + 0.07 \dfrac{(\speed+\fspeed)^2}{\speed } 
\end{equation}

\begin{figure}
\begin{center}
\hspace*{-1cm}\includegraphics[width=11cm]{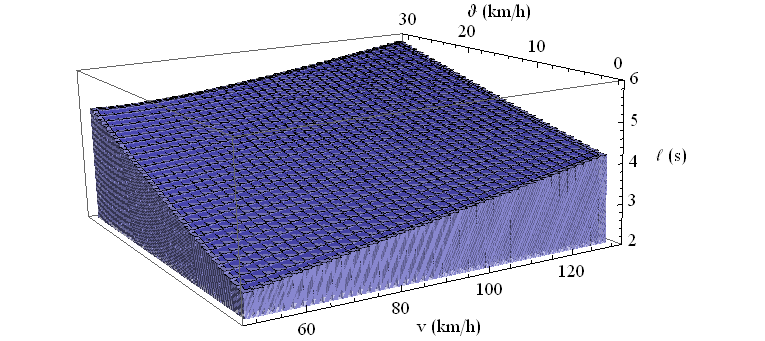} %, angle =90
\end{center}
\caption{Safety $\lsec$ headway in relation to adversarial speed manipulation $\fspeed \in [0..30]$ and vehicle reported speed $\speed \in [0..130]$}
\label{fig:speed_mspeed_adv}

\begin{center}
\hspace*{-1cm}\includegraphics[width=11cm]{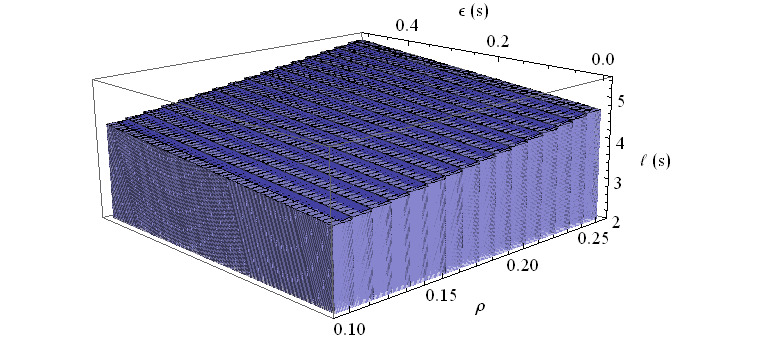} %, angle =90
\end{center}
\caption{Safety $\lsec$ headway in relation to adversarial speed manipulation $\ratio\speed$ and delay in reaction time $\freact \in (0..0.5)$ at reported vehicle speed $\speed=90km/h$}
\label{fig:speed_ratio_adv}
\end{figure}

We discuss the impact of this on graphical representations from 3D plots. 
Figure \ref{fig:speed_mspeed_adv} depicts the safety $\lsec$-seconds headway in relation to adversarial speed manipulation $\fspeed \in [0..30]$ and vehicle speed $\speed \in [0..130]$. The safe headway is between 3 and 6 seconds.
Figure \ref{fig:speed_ratio_adv} depicts the safety $\lsec$-seconds headway in relation to adversarial speed manipulation as ratio from the actual speed $\ratio\speed$ (10--25\% considered) and the delay in reaction time $\freact \in (0..0.5)$ at reported vehicle speed $\speed=90km/h$. Similar to Figure \ref{fig:speed_mspeed_adv}, adversarial manipulation increases the safety margin to 4--6 seconds, but note that in contrast to Figure \ref{fig:speed_mspeed_adv} now the reported vehicle speed is bound to only $\speed=90km/h$.
Roughly speaking, assuming adversarial manipulation of taillights, with a driver reaction time delayed to at least 2 seconds, consistent with the report in \cite{Lee76}, we have:

\begin{equation}
 \lsec > 2 +2\ratio + 0.019 (\ratio+1)^2 \speed 
\end{equation}
 
 This would dictate a safety rule of at least \emph{3 seconds plus 1 second for each 25 km/h}.

\newcommand\lland{\mathsf{~and~}}
\newcommand\llor{\mathsf{~or~}}

\section{Models for simulation and results}

In this section we derive models that are suitable for the simulation of vehicle collisions on map overlays. We start with a simple model for a single lane and  continue with a multiple lane intersection. For both we derive experimental results in order to garner some sense of reality and of the impact on real-world intersections.

\subsection{Model for a vehicle formation on a single lane}

A vehicle formation on a single lane heading toward an obstacle is suggested in Figure \ref{fig:veh_single_lane} (the image is an overlay over a map selected at random from OpenStreetMap\footnote{www.openstreetmap.org/}). The obstacle is instanced in our scenario by a traffic light. Using the traffic light as an obstacle is not accidental as this object is common part of the environment and it can be also manipulated by an adversary. In a worst case scenario, vehicles heading toward it can have their speed modified and the traffic light may be delayed, answering clearly to the theoretical scenarios discussed in the previous section.

To derive collisions, modelling vehicle speed is necessary. 
Vehicle speed is easy to adjust by considering the states of the vehicle: i) the initial state when the vehicle is running at $\vinit$, ii) the braking stage and iii) the point when the vehicle stops or collides with another vehicle or reaches the obstacle.
The vehicle is crashed and the speed is 0 when the distance to the vehicle in front (or the obstacle) is smaller than the vehicle length. Distinct to the theoretical models in section II, we also embed here the length of the vehicles in defining a collision. This is more realistic for a practical model as two vehicles need a headway of one vehicle or they collide, but has a smaller relevance from a gross theoretical estimation as expressed in Section II. Until the driver reacts, i.e., time $i \dtime$ for the i-th driver, the speed remains $\vinit$. From the time at which the driver starts braking, i.e., $t \geq i  \dtime$, the speed decreases by $(t-i \dtime)\mu  g$. 

The following equation incorporates speed modifications and the position of the vehicle which is adjusted based on speed at a simulation step $\Delta t$:

\begin{equation}
\begin{cases}
v_i(t)=
\begin{cases}
               0 \ifff \dist{x_i}{x_{i-1}} < \vlen \\
               \vinit \ifff  \dist{x_i}{x_{i-1}} \geq \vlen \lland t < i \dtime\\
			   \begin{aligned} 
			   &\vinit-(t-i \dtime)  \mu  g \ifff\\ 
			   &~~~~~~~~~~~\dist{x_i}{x_{i-1}} \geq \vlen \lland t \geq i  \dtime
			   \end{aligned} 
\end{cases} \\
x_i(t+\Delta t) = x_i(t)+v_i(t)  \Delta t
\end{cases} 
\end{equation}

\begin{figure}
\begin{center}
\includegraphics[width=8.5cm]{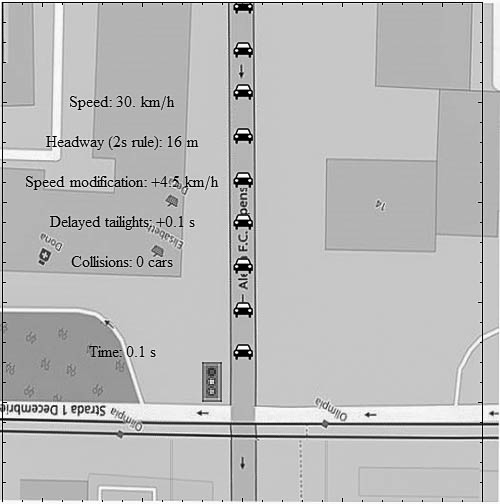} 
\end{center}
\caption{Simulation of vehicles on a single lane heading toward obstacle as map overlay: 1 collision at 4.7s (left) and 6 collisions at 16.1s (right)}
\label{fig:veh_single_lane}
\end{figure}

In Table \ref{tab:coll_fs} we show the number of collisions as derived from our simulation. We account for vehicle speeds $v \in \{20, 30, 50, 90\}$ (km/h), adversarial modifications $\fspeed$ at 5\% or 15\% of the original speed and delayed reaction time by $100 ms$ or $200 ms$. The number of collisions is shown which starts from a single vehicle at $v=20 km/h$ and a modification of just $1 km/h$ with a $100 ms$ delay for the traffic light. At $v=90 km/h$ and a modification of just $13.5 km/h$ with a $200 ms$ delay the number of collisions is $\ccount = 53$ vehicles. Our model confirms the value of $\ccount$ which also follows directly from Equation \ref{eq:count}. Care should be taken at choosing $\Delta t$ since at higher speeds even a smaller $\Delta t$ can lead to significant loss in the accuracy of the results. We generally worked in our simulations with delays from a dozen to several hundred milliseconds, the smaller the delay the higher the accuracy. 

\begin{table*}[h]
\begin{center}
\caption{Collisions on a single lane with adversarial modified speed $\fspeed$ and delayed reaction time $\freact$}
\label{tab:coll_fs}
  \begin{tabular}{|c  |c| c |c| c | c| c| c| c | c |c| c| c | c |c| c| c| }
    \hline     
 speed (km/h) & \multicolumn{4}{c|}{20} & \multicolumn{4}{c|}{30} & \multicolumn{4}{c|}{50} & \multicolumn{4}{c|}{90}  \\ 
 \hline
     $b$ (2-s rule)&  \multicolumn{4}{c|}{11} & \multicolumn{4}{c|}{16}   & \multicolumn{4}{c|}{27}     & \multicolumn{4}{c|}{50}    \\ 
 \hline %\hline
    $\fspeed$ (km/h) 			& \multicolumn{2}{c|}{1} & \multicolumn{2}{c|}{3} & \multicolumn{2}{c|}{1.5} & \multicolumn{2}{c|}{4.5}  &  \multicolumn{2}{c|}{2.5} & \multicolumn{2}{c|}{7.5}  & \multicolumn{2}{c|}{4.5} & \multicolumn{2}{c|}{13.5}     \\ 
\hline
	$\freact$ (ms) 				& 100 	& 200  	& 	100  	& 200  	& 100 	& 200 	& 100  	& 200  	& 100  	& 200  	& 100  	& 200  	& 100  	& 200 	& 100  	& 200\\ 
\hline \hline
	\rowcolor{lightgray} $\ccount$ (collided cars) 	&  1		&   2		&  	3		&  11	 	&  2  	&    3		&   5		&  17		&  3	&   	5	&   8		&  29		&  	6	&   9		&   15		&  53 \\ 
\hline     \hline
  \end{tabular}
 \end{center}  
\end{table*}

Figure \ref{fig:dist_sim} (left) gives the distances between the cars and the obstacle at $v=20 km/h$, $\fspeed = 3 km/h$, $\freact=200ms$, a case for which 11 cars collided (see Table \ref{tab:coll_fs}). For the simulation we considered 20 vehicles, then set $\Delta t = 10 ms$ and run $4000$ steps, a point at which all vehicles stopped. Distances between cars appear to be equal but on a closer look to Figure \ref{fig:dist_sim}  the first 11 cars have a headway of less than 5 meters while the next 9 cars have a headway of only slightly more than 5 meters. Thus the last 9 cars were extremely close to a collision as well. Figure \ref{fig:speed_sim} shows the speed evolution (left) and distance between cars and stoplight (right) in 4000 simulation steps for the same speed and adversarial modifications as previous.

\begin{figure}
\begin{minipage}{4.25cm}
\includegraphics[width=4.25cm]{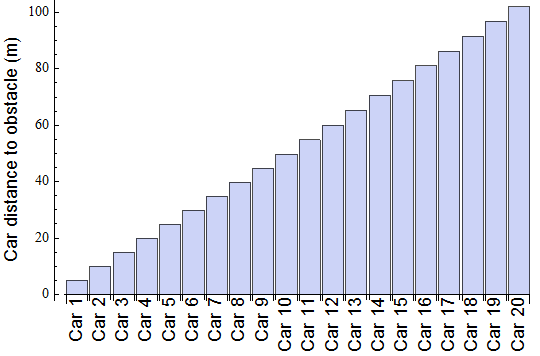} %, angle =90
\end{minipage}
\begin{minipage}{4.25cm}
\includegraphics[width=4.25cm]{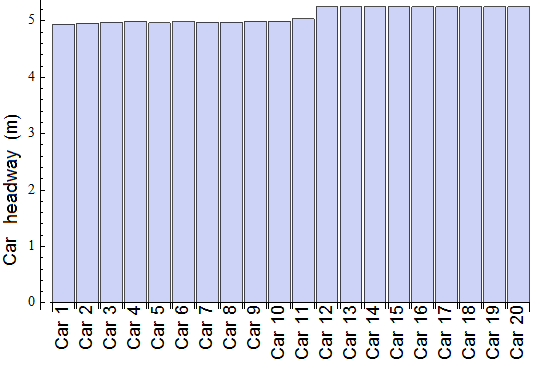} %, angle =90
\end{minipage}
\caption{Distance between each cars and the traffic light (left) and distance between cars (right) at $v=20 km/h$, $\fspeed = 3 km/h$, $\freact=200ms$}
\label{fig:dist_sim}

\begin{minipage}{4.25cm}
\includegraphics[width=4.25cm]{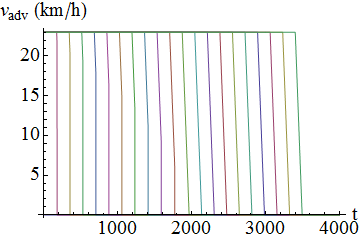} %, angle =90
\end{minipage}
\begin{minipage}{4.25cm}
\includegraphics[width=4.25cm]{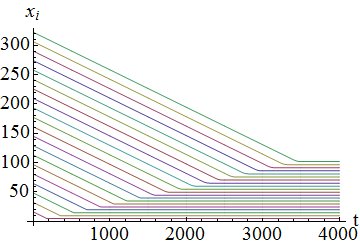} %, angle =90
\end{minipage}
\caption{Speed evolution (left) and distance between cars and stoplight (right) in 4000 simulation steps at $\speed=130km/h$, $b=72 m$ at $v=20 km/h$, $\fspeed = 3 km/h$, $\freact=200ms$}
\label{fig:speed_sim}
\end{figure}

\begin{figure}[h!]
\centering
\begin{minipage}{4.25cm}
%\begin{center}
\includegraphics[width=4cm]{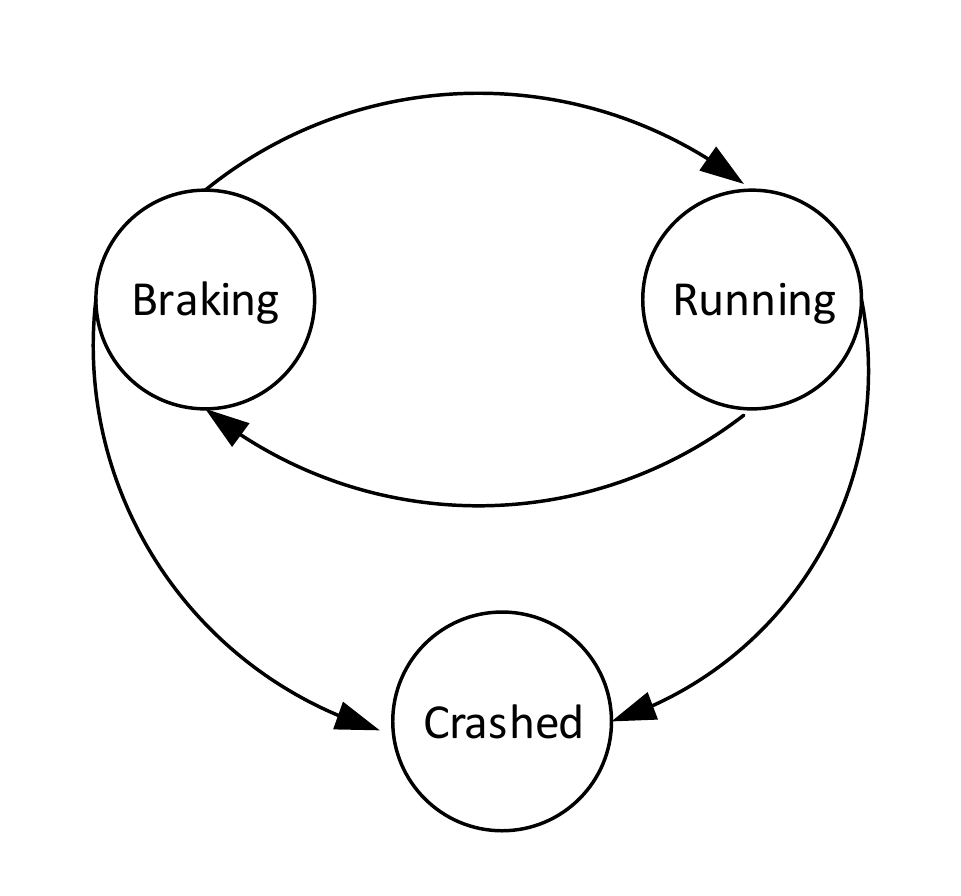}
%\end{center}
\end{minipage}
\caption{State transitions for a vehicle}
\label{fig:veh_state}
\end{figure}

\subsection{Modelling multiple vehicle collisions at a crossroad}

We now move to a more complex and more realistic scenario: a vehicle crossroad as depicted in Figure ~\ref{fig:veh_intersection}. This image is created as an overlap of our simulation on a real-world intersection but names on the map are removed since the scenario here is imaginary. The real-world intersection was selected mostly at random from OpenStreetMap only to serve as an example and we are not aware of specific traffic details. A traffic simulation that is fully accurate to the real-world model is not part of our goals here but may be subject of future work and is easy to derive from the formalism that we introduce. Now we simply place the cars on lanes as we feel natural. In particular we consider 6 lanes with 10 cars on each, resulting in 60 cars heading toward the intersection. Vehicle size on the map is increased compared to the rest of the objects to make vehicles visible. The headway between vehicles is the 2-seconds headway and its length is proportional with the size of the car. 

\begin{figure*}
\begin{center}
\includegraphics[width=17cm]{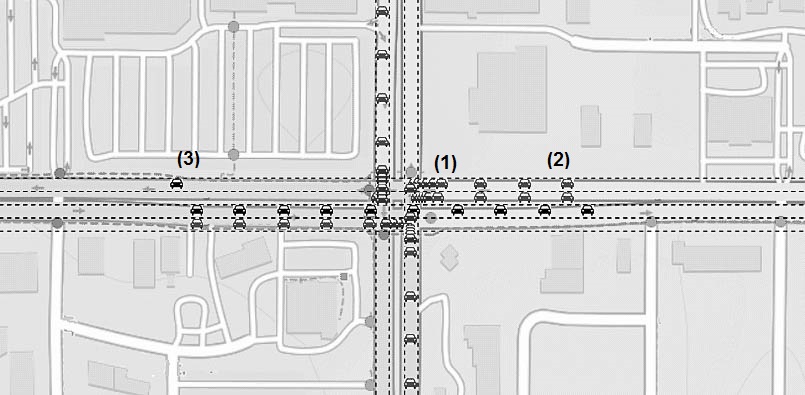} %, angle =90
\end{center}
\caption{Simulation of vehicles on multiple lanes heading toward crosspoint as overlay on a map: (1) vehicle collision in the middle of the intersection, (2) vehicles that are braking and (3) one vehicle departing from the intersection}
\label{fig:veh_intersection}
\end{figure*}

Modelling requires slight improvements over the previous equations.
We need to refine some notions by giving more comprehensive definitions for vehicle and lanes, etc. In our simulation we used the following formalism:

\begin{enumerate}
\item a \textbf{vehicle} is represented as a structure containing four elements: speed $\speed \in [0, \infty ]$, position $\pos \in [-\infty, \infty] $, state $\stt \in \{ \running, \braking, \crashed \}$ and the time at which state $\braking$ was reached $\tstamp \in [0, \infty ]$, i.e., $\vehicle= \{ \speed, \pos, \stt, \tstamp \}$,

\item a \textbf{vehicle formation} $\squadron$ is a collection of vehicles, i.e., $\squadron =\{ \vehicle_0, \vehicle_1, ..., \vehicle_{n-1} \}$,

\item a \textbf{lane} is represented as a structure containing four elements: start-point $\startp$, end-point $\stopp \in \real \times \real$, direction $\direction \in [-1, 1]$, angle $\angl \in [0, 2\pi]$ and stop signs $\stopsigs=\{d_0, d_1, ..., d_{l-1}\}$ (where $d_i, i=0..l-1$ denotes the position of each stop sign), i.e., $\lane = \{ \startp, \stopp, \direction, \angl, \stopsigs \}$,

\item we define a \textbf{traffic model} $\tmodel$ as a collection of lanes $\lanes = \{ \lane_0, lane_1, ..., \lane_{n-1}\}$ each holding one vehicle formation $\squadron_i, i \in \{0..n-1\}$, i.e., $\tmodel = \lanes \times \squadron $,

\item the \emph{intersection points} of a \textbf{traffic model} $\tmodel$  are the list of pairs $\intersection = \left \{ p_0=\{(x'_0, y'_0),(x''_0, y''_0)\} ... p_l=\{(x'_l, y'_l),(x''_l, y''_l)\}\right \}$.

\end{enumerate}

Again each vehicle must start braking either when the vehicle in front brakes or when the stop sign becomes visible. We find it easier to visualize the vehicle as transiting between the three states: running, braking or crashed as depicted in Figure \ref{fig:veh_state}. A car is crashed if is already crashed or there exists another vehicle that collides with it in the current step. If it is not crashed then the car is running if it is not braking and is braking if the vehicle in front does so or the stop sign becomes visible. This is summarized by the following formalism for the vehicle state:

\begin{multline}
\small
\begin{cases}
				\begin{aligned} 
               \crashed \ifff &\iscrashed(\vehicle_{i}) \lor \exists k.\collision(\vehicle_{i}, \vehicle_k)  
				\end{aligned} \\
			   \begin{aligned} 
			   \running \ifff \neg \iscrashed(\vehicle_{i}) \land  \neg \isbraking(\vehicle_{i}) 
			   \end{aligned} \\
               \braking \ifff  \neg \iscrashed(\vehicle_{i}) \\ ~~~~~~~~~~~~~~~\land \bigl ( \isbrakingorcrashed(\vehicle_{i-1}) \lor \stopvis(\vehicle_i) \bigr ) 
			   \end{cases} \nonumber
\end{multline}

We use several predicates to get the state of a vehicle, i.e., $\isrunning(\vehicle_{i})$, $\isbraking(\vehicle_{i})$,  $\iscrashed(\vehicle_{i})$,  $\isbrakingorcrashed(\vehicle_{i})$, to determine collisions between vehicles, i.e., $\collision(\vehicle_{i}, \vehicle_j)$,  and to establish if a stop-sign is visible for a car, i.e., $\stopvis(\vehicle_{i})$. These can be all simply derived from the car location on the map.
The coordinates of each car can be easily extracted from the position of the car on the lane, the angle of the lane and its coordinates as:

\begin{equation}
\getx(\vehicle_{i}) = x_\lane \cdot \sin(\angl) + \cos(\angl) \cdot \getpos(\vehicle_i)
\end{equation}

\begin{equation}
\gety(\vehicle_{i}) = y_\lane \cdot  \cos(\angl) + \sin(\angl) \cdot \getpos(\vehicle_i)
\end{equation}

Subsequently, the distance between the car and the other object can be computed as Euclidean distance. 
Checking that a car collides with another car or that a stop-sign is visible simply requires checking the distance between objects. 
Two vehicles collide if the distance between them is smaller than the vehicle length and a stop sign becomes visible as soon as it reaches the visual range of the driver.

To run the simulation we need rules for updating vehicle speed $\speed$, position $\pos$ and state $\stt$. The state $\stt$ is updated as shown in Figure \ref{fig:veh_state}, the vehicle runs if the car in front is not crashed and not  braking, otherwise the vehicle brakes. Similarly, vehicles brake if the stop-sign is visible. Speed adjustment is done according to the vehicle state and the previously defined adjustment rules, the same is done for vehicle position:

\begin{equation}
\begin{cases}
v_i(t)=
\begin{cases}
               0 \ifff \iscrashed(\vehicle_{i}) \\
               \vinit \ifff \isrunning(\vehicle_{i}) \\
			   \vinit- (t- \tstamp)  \mu  g \ifff  \isbraking(\vehicle_{i}) \\
\end{cases} \\
x_i(t+\Delta t) = x_i(t)+v_i(t) \Delta t
\end{cases} 
\end{equation}

We now show simulation results and discuss them on graphical representations. First we consider the case of: $\speed=30km/h$, $b=16 m$, $\fspeed = 1.5 km/h$, $\freact=0ms$, i.e., no delay in the taillights, and 10 vehicle on each of the lanes from Figure \ref{fig:veh_intersection} leading to a total of 60 vehicles.
Figure \ref{fig:bar_dist_1} plots the distance of each car to the center of the intersection also showing the state of each car. Figure \ref{fig:speed_sim_1} plots the evolution for the speed of each car.
Figure \ref{fig:dist_sim_1} plots the distance of each car to the center of the intersection. Section (1) of the plot depicts the cars that are crashed or successfully brake, section (2) cars approaching the intersection and section (3) the cars that are departing from the intersection (no collision in front). 
In Figure \ref{fig:hways_sim_1} the evolution of distance between each cars and the car from the rear, i.e., the headway, is shown. We mark by (1) the cars that crashed, (2) marks the cars that stopped at a safe distance and (3) the cars that depart from the intersection and have a constant headway. Note that in sector (3) one of the cars has an increasing headway, this is the case of the car from the first horizontal lane that successfully departs from the intersection while the rest of the cars from its lane have crashed (the car can be easily identified in Figure \ref{fig:veh_intersection}).

We give similar graphical depictions for $\speed=30km/h$, $b=16 m$, $\fspeed = 4.5 km/h$, $\freact=200ms$ in Figures \ref{fig:bar_dist_2}, \ref{fig:speed_sim_2}, \ref{fig:dist_sim_2}and  \ref{fig:hway_sim_2}. The number of crashed vehicles is much higher with only 5 vehicles that successfully stopped. Again, 1 vehicle on the first lane and the 10 vehicles on the 3-rd are escaping the collision, but distinct to the previous case where more than 30 cars managed to brake, now only 5 managed to brake in time and the rest are crashed. Similarly, in Figures \ref{fig:dist_sim_2} and \ref{fig:hway_sim_2} we mark the three areas (1) crashed vehicles, (2) stopped vehicles and (3) vehicles running. In this case area (1) clearly conglomerates more crashed vehicles.

\newcommand\figw{6.75}

\begin{figure*}
\begin{minipage}{9cm}
\includegraphics[width=9cm]{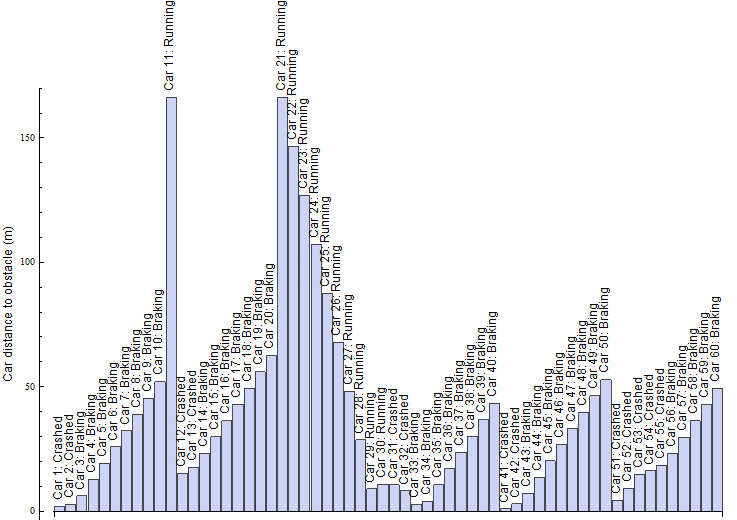} %, angle =90
\caption{Distance between each cars at $\speed=30km/h$, $b=16 m$, $\fspeed = 1.5 km/h$, $\freact=0ms$}
\label{fig:bar_dist_1}
\end{minipage}
\begin{minipage}{9cm}
\includegraphics[width=9cm]{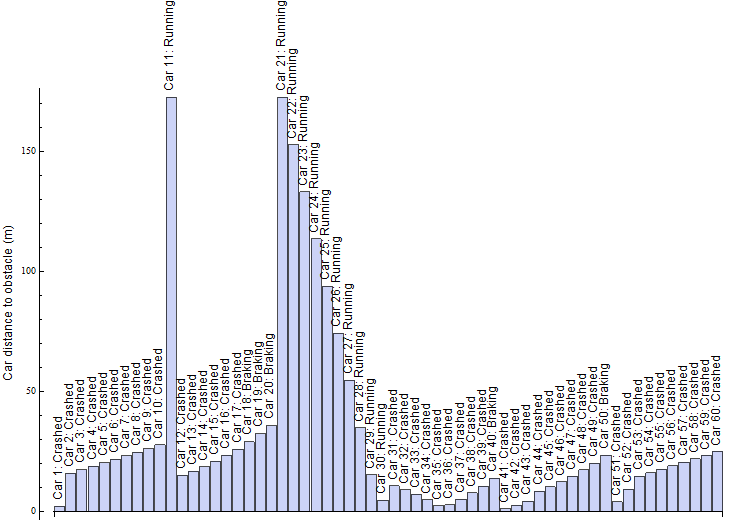} %, angle =90
\caption{Distance between each cars at $\speed=30km/h$, $b=16 m$, $\fspeed = 4.5 km/h$, $\freact=200ms$}
\label{fig:bar_dist_2}
\end{minipage}
%\end{figure*}
%\begin{figure*}[tb!]
%\begin{figure}[t!]
\begin{minipage}{9cm}
\centering
\includegraphics[width=\figw cm]{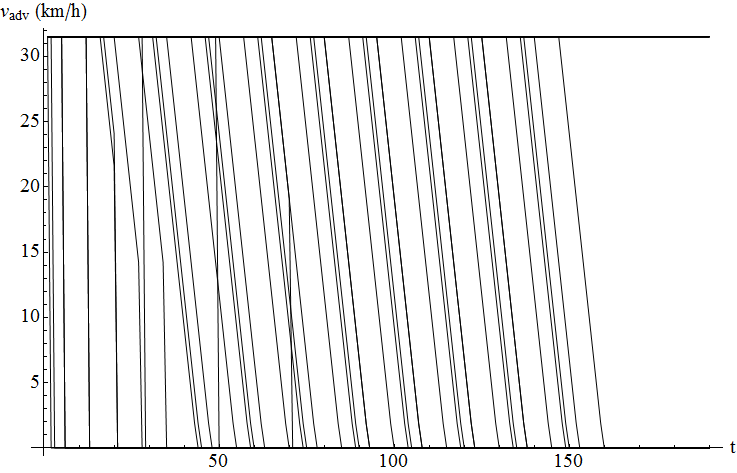} %, angle =90
\caption{Speed evolution at $\speed=30km/h$, $b=16 m$, $\fspeed = 1.5 km/h$, $\freact=0ms$}
\label{fig:speed_sim_1}
~~\\
~~\\
\centering
\includegraphics[width=\figw cm]{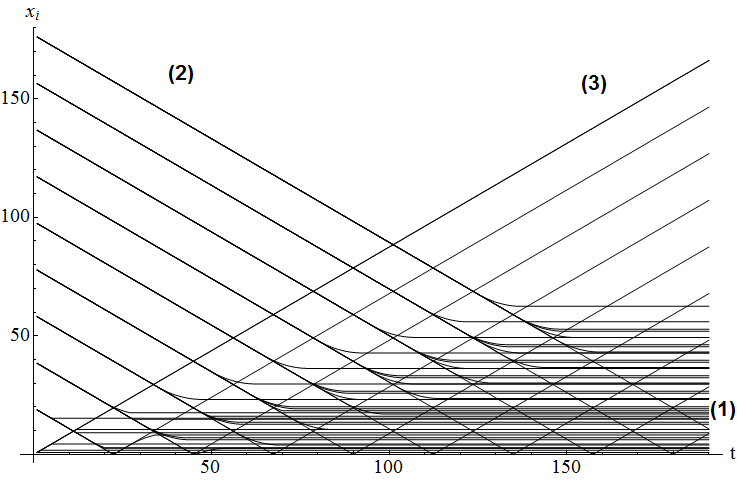} %, angle =90
\caption{Distance evolution at $\speed=30km/h$, $b=16 m$, $\fspeed = 1.5 km/h$, $\freact=0ms$}
\label{fig:dist_sim_1}
~~\\
~~\\
\centering
\includegraphics[width=\figw cm]{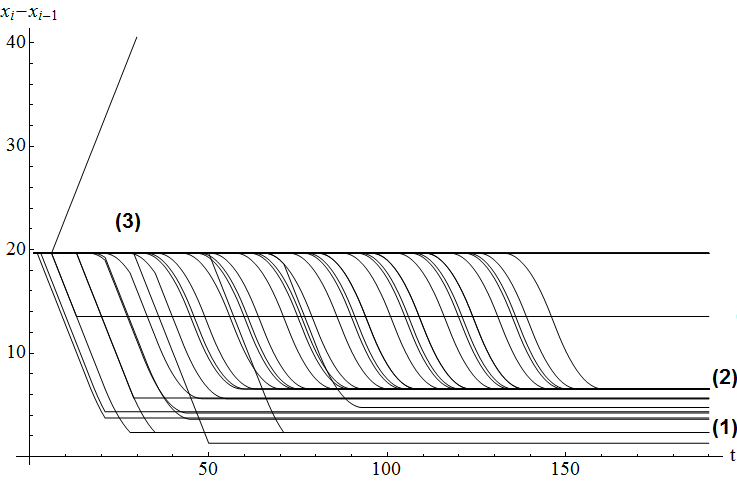} %, angle =90
\caption{Headway evolution at $\speed=30km/h$, $b=16 m$, $\fspeed = 1.5 km/h$, $\freact=0ms$}
\label{fig:hways_sim_1}
\end{minipage}
%\end{figure}
\begin{minipage}{9cm}
%\begin{figure}[t!]
\centering
\includegraphics[width=\figw cm]{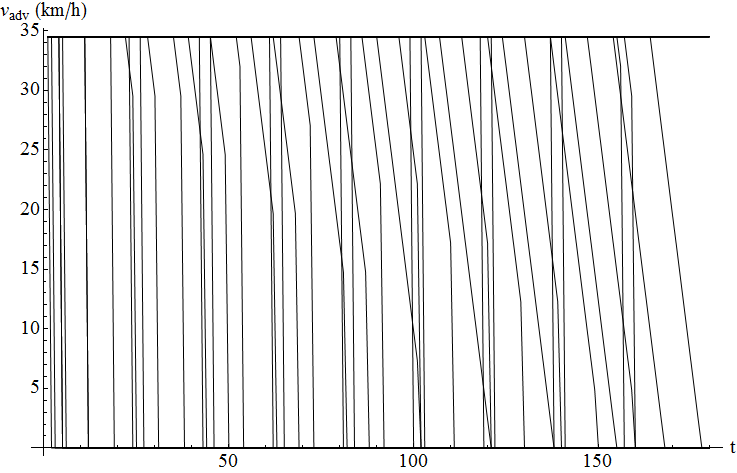} %, angle =90
\caption{Speed evolution at $\speed=30km/h$, $b=16 m$, $\fspeed = 4.5 km/h$, $\freact=200ms$}
\label{fig:speed_sim_2}
\centering
\includegraphics[width=\figw cm]{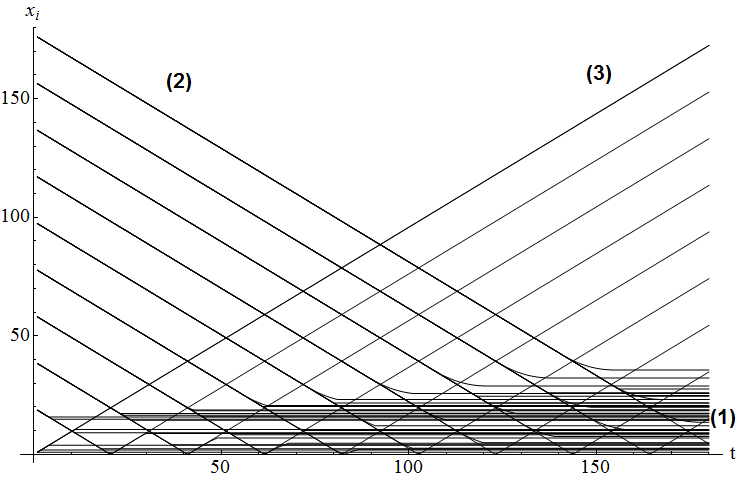} %, angle =90
\caption{Distance evolution at $\speed=30km/h$, $b=16 m$, $\fspeed = 4.5 km/h$, $\freact=200ms$}
\label{fig:dist_sim_2}
~~\\
~~\\
\centering
\includegraphics[width=\figw cm]{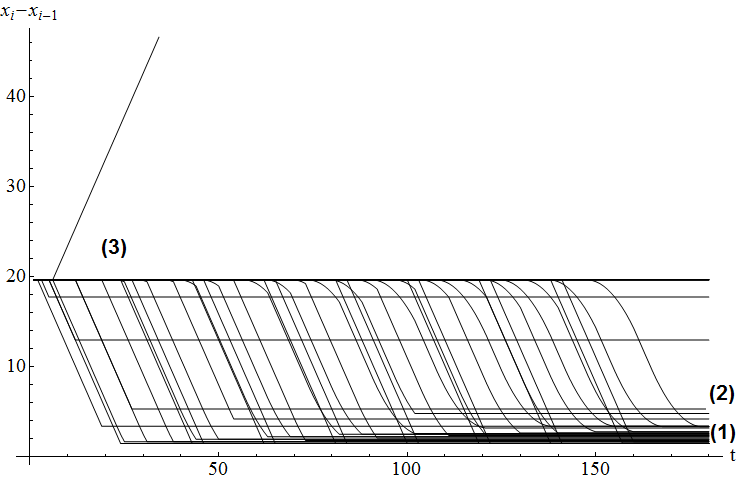} %, angle =90
\caption{Headway evolution at $\speed=30km/h$, $b=16 m$, $\fspeed = 4.5 km/h$, $\freact=200ms$}
\label{fig:hway_sim_2}
%\end{figure}
\end{minipage}
\end{figure*}

\newcommand\thoriz{\theta}
\newcommand\totalspeed{\fspeed_{\mathit{total}}}
\newcommand\lgain{\chi}
\newcommand\const{\wp}
\newcommand\sfact{\sigma}
\newcommand\allfspeeds{\Psi}
\newcommand\allone{\mathrm{J}}
\newcommand\allpos{\mathrm{X}}
\newcommand\rpf[2]{\mathsf{AdvForm}\bigl[#1, #2\bigr]}
\newcommand\noveh{k}
\newcommand\rsquadron{adversarial platoon formation~}
\newcommand\hway{b}
\newcommand\rate{\rho}
\newcommand\prob[1]{\mathsf{Pr}\biggl\{ #1 \biggr \}}
\newcommand\rnoveh{\widetilde{\mathsf{X}}}
\newcommand\vehrate{\alpha}
\newcommand\prcor{p_{\mathit{adv}}}

\section{Intelligent adversarial behavior}

We design intelligent adversarial behaviour around two actions: \emph{\rsquadron} and \emph{stealthy speed modifications}. By the first we account for the adversary ability to coagulate a formation of cars for which it manipulates their speeds. By the former we account for speed modifications that are smooth and harder to detect by human agents.

\begin{figure*}
\begin{center}
\includegraphics[width=18cm]{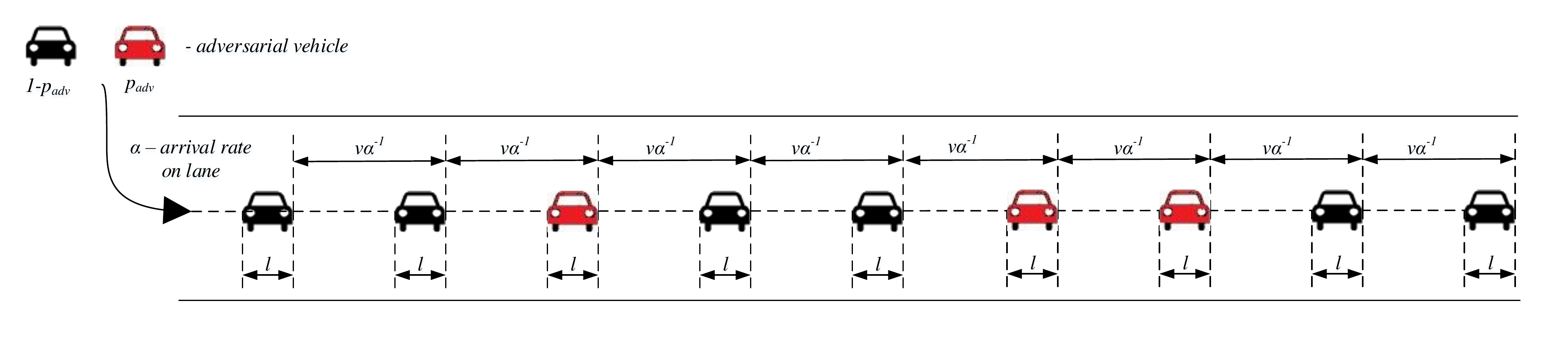} %, angle =90
\end{center}
\caption{Setup for adversarial platoon formation: vehicles arriving on the lane at rate $\vehrate$  and with corruption probability $\prcor$}
\label{fig:rogue}
\end{figure*}

\subsection{Adversarial platoon formation}

The addressed setup is suggested in Figure \ref{fig:rogue}. Vehicles are depicted arriving on the lane at a constant rate $\vehrate$. For simplicity we assume that vehicles arrival times are equidistant, this leads to a headway $b^{*}=\speed \vehrate^{-1}$. The adversary target is to coagulate compromised cars in a single platoon at headway $\hway$.
We quantify adversarial capability for \rsquadron in a theorem that fixes the probability for an adversary to form a platoon of some fixed size in a given time $T$. We then instantiate this result with practical values to give some hints on adversarial capabilities.

\begin{theorem}
\textit
Let a vehicle lane and the following predefined constants: the imposed vehicle speed on the lane $\speed$, the legal headway between vehicles $\hway$, the  arrival rate of the vehicles on the lane $\vehrate$, the probability that a vehicle is corrupted by the adversary $\prcor$ and the maximum modification rate $\rate$ of speed by adversary intervention. Assume that the time to cover the entire length of the lane at speed $\speed$ is longer than some fixed value $T$ (this fixes the time-horizon for adversarial actions). Then there exists an adversary capable to form platoons of expected size $\prcor N $ where:

\begin{equation}
N =  \dfrac{2\rate \speed T  + \hway}{ \speed \vehrate^{-1} + \hway}
\end{equation}

Moreover, let:

$$
\zeta_{k, T} = \prob{\rpf{k}{T} }
$$

the probability of the event $\rpf{k}{T}$ that the adversary constructs an \rsquadron of exactly $k$ cars in time $T$. Then:

\begin{equation}
\zeta_{k, T} = \dfrac{N!}{k!(N-k)!}\prcor^k(1-\prcor)^{N-k}
\label{eq:prob}
\end{equation}

and in case of small corruption rates $\prcor$ and large time horizon $T$, by Poisson approximation:

\begin{equation}
\zeta_{k, T} \approx e^{-N\prcor} \dfrac{(N\prcor)^k}{k!}
\label{eq:prob_poiss}
\end{equation}.

\end{theorem}

\emph{Proof.}
We consider a discrete time simulation with the length of each step set at $\Delta t$. For the fixed time horizon $T$ and simulation step $\Delta t$, let the number of steps be $\thoriz=T/\Delta t$. We define the speed manipulation for each vehicle in each time-step $\Delta t $ during time horizon $\thoriz$ as:

\begin{equation}
\allfspeeds_{\noveh,\thoriz}=
\begin{bmatrix}
    \fspeed_{1}^1 & \fspeed_{2}^1 & \fspeed_{3}^1 & \dots  & \fspeed_{\thoriz}^{1} \\
    \fspeed_{1}^2 & \fspeed_{2}^2 & \fspeed_{3}^2 & \dots  & \fspeed_{\thoriz}^{2} \\
    \vdots & \vdots & \vdots & \ddots & \vdots \\
    \fspeed_{1}^{\noveh}& \fspeed_{2}^{\noveh} & \fspeed_{3}^{\noveh} & \dots  & \fspeed_{\thoriz}^{\noveh} \\
\end{bmatrix} \nonumber
\end{equation}

and the initial positions of the vehicles as:

\begin{equation}
\allpos_{\noveh,1}(0)=
\begin{bmatrix}
    x_1(0)  \\
    x_2(0)  \\
    \vdots \\
    x_\noveh(0) \\
\end{bmatrix} \nonumber
\end{equation}

We define the all-ones matrices:

\begin{equation}
\allone_{\noveh,\thoriz}=
\begin{bmatrix}
    1 & 1 & 1 & \dots  & 1 \\
    1 & 1 & 1 & \dots  & 1 \\
    \vdots & \vdots & \vdots & \ddots & \vdots \\
    1 & 1 & 1 & \dots  & 1 \\
\end{bmatrix},
\allone_{\noveh,1}=
\begin{bmatrix}
    1  \\
    1  \\
    \vdots \\
    1 \\
\end{bmatrix} \nonumber
\end{equation}

Then assuming constant vehicle speed $\speed$, at time $\thoriz$ the positions of the vehicles is given by:

\begin{equation}
\allpos_{\noveh,1}(\thoriz) = \biggl ( \speed \cdot \allone_{\noveh,\thoriz} + \allfspeeds_{\noveh,\thoriz} \biggr) \times \biggl ( \Delta t \cdot \allone_{\noveh,1} \biggr )+  \allpos_{\noveh,1}(0)   \nonumber
\end{equation}

Where the center dot $\cdot$ denotes scalar multiplication and the multiplication sign $\times$  denotes vector product. To coagulate all vehicles in a single platoon, we need the headway between vehicles after time $\thoriz$ to be equal to constant $\hway$ regardless of initial positions given in $\allpos_{\noveh,1}(0)$. Concretely, we have:

\begin{equation}
x_i(\thoriz) - x_{i-1}(\thoriz) = \hway, \forall i=2..\noveh \nonumber
\end{equation}

This is equivalent to:

\begin{equation}
\begin{cases}
               \Delta t \sum_{i=1,\thoriz} \biggl ( \fspeed_i^2 - \fspeed_i^1 \biggr ) + x_2(0) - x_1(0) = \hway  \\
               \Delta t \sum_{i=1,\thoriz} \biggl ( \fspeed_i^3 - \fspeed_i^2 \biggr ) + x_3(0) - x_2(0) = \hway  \\
			   \dots \\
			   \Delta t \sum_{i=1,\thoriz} \biggl ( \fspeed_i^{\noveh} - \fspeed_i^{\noveh-1} \biggr ) + x_{\noveh}(0) - x_{\noveh-1}(0) = \hway  \\
\end{cases} \nonumber
\end{equation}

By summing up all of the above lines we get:

\begin{multline}
               \Delta t \sum_{i=1,\thoriz} \biggl ( \fspeed_i^{\noveh}  - \fspeed_i^1 \biggr ) + x_{\noveh}(0) - x_1(0) = \hway (\noveh-1) \\ \nonumber
			   \Rightarrow \sum_{i=1,\thoriz} \biggl ( \fspeed_i^{\noveh}  - \fspeed_i^1 \biggr ) = \dfrac{\hway (\noveh-1) - x_{\noveh}(0) + x_1(0)}{\Delta t }
\end{multline}

Note that this relation is independent of the target vehicle speed $\speed$ since all drivers intend to maintain it and cancels upon summation. Consequently, speed manipulation must compensate for the headway of $\noveh$ vehicles, i.e., $\hway (\noveh-1)$, and the difference between the initial positions of the two vehicles, i.e., $ x_{\noveh}(0) - x_1(0)$.

If there exists adversarial manipulation vectors $\fspeed^{\noveh}$ and $\fspeed^1$ such that  previous relation holds for time horizon $\thoriz$, then for all other cars there exists $\fspeed^{j}, j=2..\noveh-1$ to satisfy this relation. This is because the distance between them and the lead car is smaller and can be recovered with lesser speed manipulations.

Assume now the maximum adversarial speed manipulation as rate $\rate$ of the actual speed, i.e., $\fspeed = \rate \speed$. Worst case, the speed of vehicle $\noveh$ needs to be modified by $\rate\speed$ and that of vehicle $1$ by $-\rate\speed$ (the lead vehicle must go slower for the other to recover distance). Thus the condition for the vehicles to reach the \rsquadron formation is satisfied if and only if:

\begin{equation}
2\thoriz \rate \speed \geq  \dfrac{\hway (\noveh-1) - x_{\noveh}(0) + x_1(0)}{\Delta t } \nonumber
\end{equation}

From which we have:

\begin{equation}
x_1(0) - x_{\noveh}(0)  \leq 2\Delta t \thoriz \rate \speed - \hway (\noveh-1)  \nonumber
\end{equation}

Note that the term form the left side denotes the distance between the first and the last compromised car. We translate this to the time at which each car arrives on the lane. The time at which the first car arrives is $t=0$ since this is the lead car. Then let the time at which the $\noveh$-th car arrives be $t_{\noveh}$. Having arrival rate $\vehrate$ we have $t_{\noveh} = \noveh \vehrate^{-1}$ and:

\begin{equation}
x_1(0)-x_{\noveh}(0)=\speed t_{\noveh}= \speed \noveh \vehrate^{-1}
 \nonumber
\end{equation}

Which leads to:

\begin{equation}
 \noveh \leq \dfrac{2\rate \speed T  + \hway}{ \speed \vehrate^{-1} + \hway}  \nonumber
\end{equation}

This fixes the maximum number of vehicles for which adversarial behaviour can be accounted in time $T$, i.e., 

\begin{equation}
 N =  \dfrac{2\rate \speed T  + \hway}{ \speed \vehrate^{-1} + \hway}  \nonumber
\end{equation}

Now equation \ref{eq:prob} simply gives the probability of $k$ success out of $N$ in a Bernoulli trial with probability $\prcor$. A sufficiently large time horizon $T$ implies a larger $N$ and a small $\prcor$ leads to the Poisson approximation in equation \ref{eq:prob_poiss}.

\begin{figure}
\begin{center}
\includegraphics[width=7.5cm]{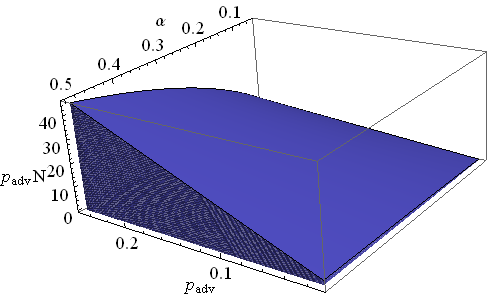} %, angle =90
\end{center}
\caption{The expected platoon size at $\speed = 130 km/h$, $T=1 h$ with $\prcor \in [0.01, 0.25]$ and arrival rate $\vehrate \in [0.05, 0.5]$}
\label{fig:expect_size}
%\end{figure}

%\begin{figure}
\begin{center}
\includegraphics[width=7.5cm]{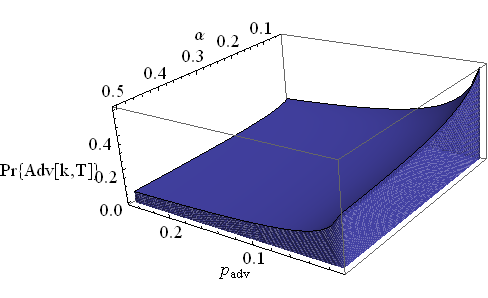} %, angle =90
\end{center}
\caption{Probability of adversarial platoons of the expected size at $\speed = 130 km/h$, $T=1 h$ with $\prcor \in [0.01, 0.25]$ and arrival rate $\vehrate \in [0.05, 0.5]$}
\label{fig:prob_expected}
%\end{figure}

%\begin{figure}
\begin{center}
\includegraphics[width=7.5cm]{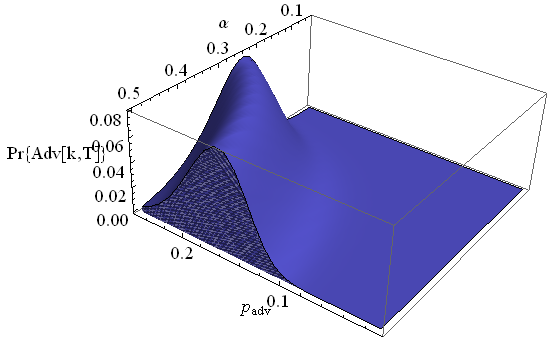} %, angle =90
\end{center}
\caption{Probability of an adversarial platoon when only 30 corrupted cars have reached the lane $\speed = 130 km/h$, $T=1 h$ with $\prcor \in [0.01, 0.25]$ and arrival rate $\vehrate \in [0.05, 0.5]$}
\label{fig:prob_30}
\end{figure}

We now consider as example the case of a lane with vehicles at speed $\speed = 130 km/h$, i.e., a high-way lane. Vehicle corruption probability is set at $\prcor \in [0.01, 0.25]$, that is, from 1 in 100 cars up to 1 in 4 cars can be adversarial. We consider arrival rate $\vehrate \in [0.05, 0.5]$, i.e., from 1 car at each 20 seconds to 1 car every 2 seconds. Figure \ref{fig:expect_size} depicts the expected platoon size under these variations. The size of the platoon can grow to almost 50 cars when corruption probability and arrival rate is high, all these cars can be concentrate by an adversary in a single platoon after 1 hour. Figure \ref{fig:prob_expected} shows probability to form a platoon of expected size which is sufficiently high, roughly between 0.15 and 0.5. Figure \ref{fig:prob_30} depicts the probability that an adversary forms a platoon of 30 cars. This probability is initially very low but steadily grows once corruption rate reaches 10\% and arrival rate grows to 1 car every 5 seconds, i.e., $\vehrate = 0.2$.

\subsection{Stealthy speed manipulation functions}

So far our models assumed constant modification of vehicle speed $\fspeed$. A sudden increase or decrease in speedometer value may however be easily noticeable by the driver. 
Research results in the area of perception clearly establish that: it is the gradualness of change that makes acceleration and deceleration difficult to perceive \cite{Schmerler75}. 
Consequently, it seems natural to turn the adversarial manipulation into a sigmoid-like function that smoothly increases and decreases over time. This seems to be consistent with  regular behavior of drivers that once starting to accelerate/brake will likely be tempted to continue further. We depict some suggestive shapes for stealthy speed modifications by an adversary in Figure \ref{fig:speed_sigmoid}.

\begin{figure}[h!]
\centering
\begin{minipage}{7cm}
\includegraphics[width=7cm]{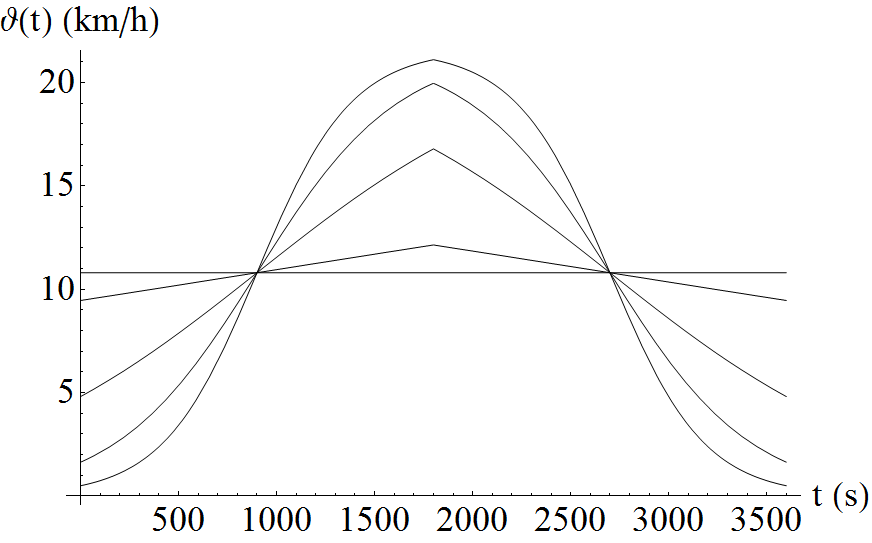}
\end{minipage}
\caption{Possible speed modifications by an adversary over a time-horizon of 60 minutes}
\label{fig:speed_sigmoid}
\end{figure}

We now extend our model to an adversary that is able to modify speeds at this finer granularity. We assume that the adversary has a fixed time horizon $T$ for achieving this goal similar to the setup provided in the previous theorem.

\begin{theorem} In the setup of Theorem 1, let the constant speed modification $\lgain$ for vehicle $k$ be:

\begin{equation}
\lgain(k)=\dfrac{x_k(0)-x_1(0)-b(k-1)}{T}
\end{equation}
\end{theorem}

Then 

\begin{equation}
\fspeed_k(t)=
\begin{cases}
               \dfrac{2\lgain(k)}{\const^{t-T/4}+1} \ifff t \in [0, T/2) \\
               2\lgain(k) \left ( 1 - \dfrac{1}{\const^{t-3T/4}+1} \right ) \ifff  t \in [T/2, T] \\
\end{cases} 
\end{equation}

Provides a smooth acceleration/deceleration adversarial modification of speed, where $\const$ is computed as function over the time-horizon $T$ and a smoothness factor $\sfact$ as

\begin{equation}
\const=1-\sfact T^{-1}
\end{equation}

\emph{Proof.} We show that the adversarial speed gain over time horizon $T$ is the same as in the case of constant speed modification, that is, we prove that:

$$
\int_{0}^{T} \fspeed(t) dt = \lgain(k) T = x_k(0)-x_1(0)-b(k-1)
$$

First note that $\fspeed(t)$ is symmetrical around $T/2$, that is:

$$
\fspeed(T/2-i) = \fspeed(T/2+i), \forall i \in [0, T/2]
$$

This follows easily since:

\begin{multline}
\fspeed(T/2+i)  = 2\lgain \left ( 1 - \dfrac{1}{\const^{T/2+i-3T/4}+1} \right ) = \\
= 2\lgain \left ( 1 - \dfrac{1}{\const^{i-T/4}+1} \right ) = 
2\lgain \left ( 1 - \dfrac{\const^{T/4-i}}{\const^{T/4-i}+1} \right ) = \\
= 2\lgain \dfrac{1}{\const^{T/4-i}+1} = \fspeed(T/2-i), \forall i \in [0, T/2] 
\end{multline}

Then:

\begin{multline}
\int_{0}^{T} \fspeed(t) dt = 2\int_{0}^{T/2} \fspeed(t) dt = 2\int_{0}^{T/2} \dfrac{2\lgain}{\const^{t-T/4}+1}  dt = \\
= 2 \biggl [ \int_{0}^{T/4} \fspeed(T/4-t)  dt + \int_{0}^{T/4} \fspeed(T/4+t) dt \biggr ] = \\
= 4 \lgain  \int_{0}^{T/4} \dfrac{1}{\const^{-t}+1}  + \dfrac{1}{\const^{t}+1}  dt = 4 \lgain  \int_{0}^{T/4} 1  dt = \lgain(k)T \nonumber
\end{multline}

which completes the proof.

\begin{figure}
\begin{minipage}{4.25cm}
\includegraphics[width=4.35cm]{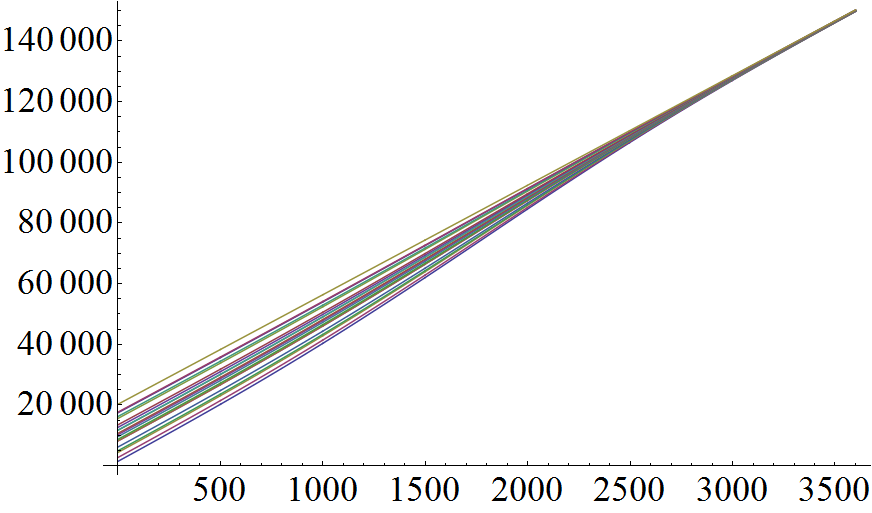}
\caption{Trajectory of the platoon during 60 minutes}
\label{fig:traj_60min}
\end{minipage}
\begin{minipage}{4.25cm}
\includegraphics[width=4.35cm]{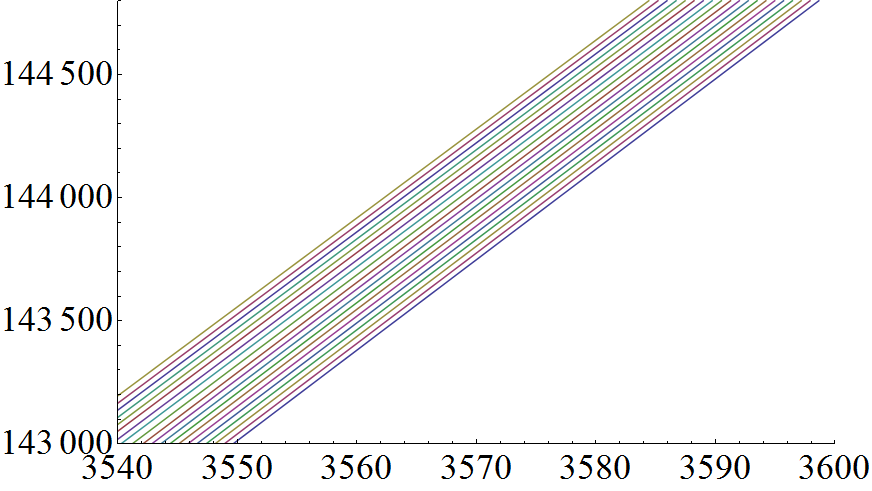}
\caption{Detail for the trajectory of the platoon during last 5 minutes}
\label{fig:traj_det_60min}
\end{minipage}
%\end{figure}
%
%\begin{figure}
\begin{minipage}{4.25cm}
\includegraphics[width=4.45cm]{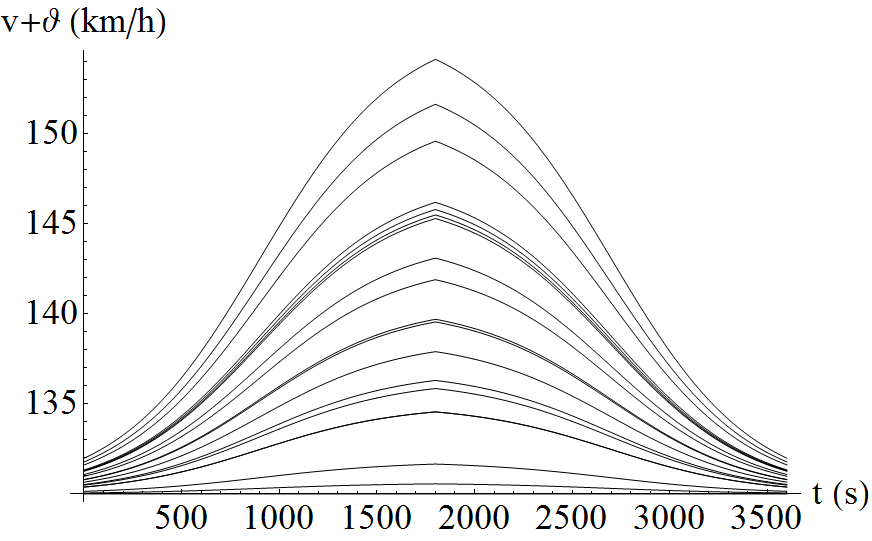}
\caption{Speed gain for each vehicle in the platoon during the 60 minutes}
\label{fig:speed_60min}
\end{minipage}
\begin{minipage}{4.25cm}
\includegraphics[width=4.45cm]{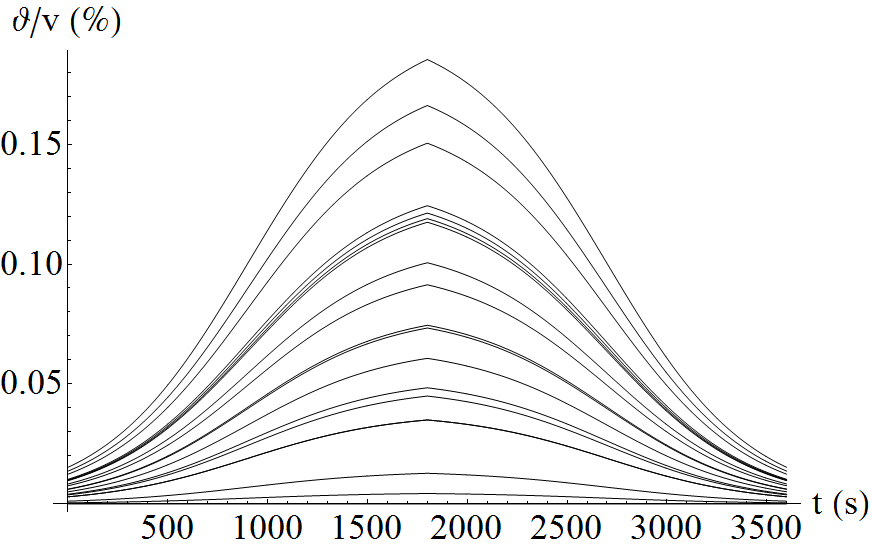}
\caption{Speed gain in percents of the reported vehicle speed during the 60 minutes}
\label{fig:speed_per_60min}
\end{minipage}
%\end{figure}
%
%\begin{figure}
\begin{minipage}{4.25cm}
\includegraphics[width=4.35cm]{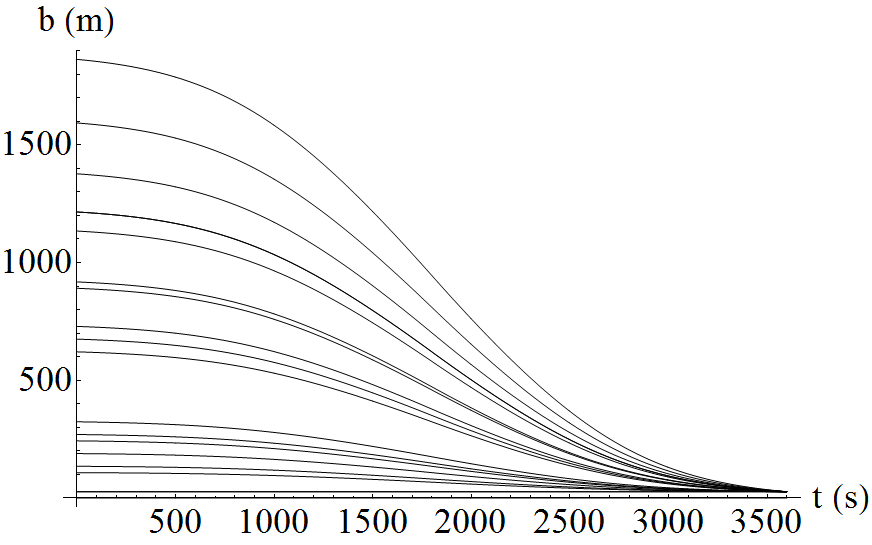}
\caption{Headway for each vehicle in the platoon during the 60 minutes}
\label{fig:hway_60min}
\end{minipage}
\begin{minipage}{4.35cm}
\includegraphics[width=4.25cm]{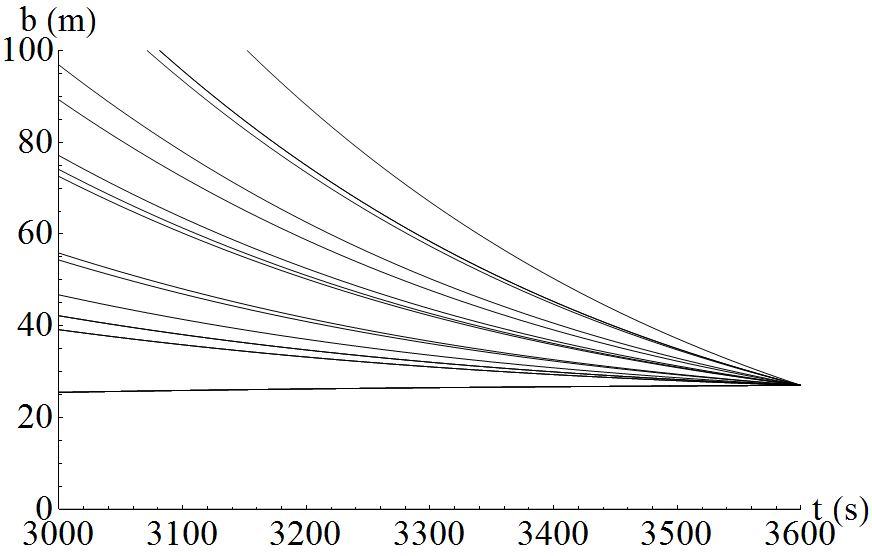}
\caption{Detail on headway for each vehicle in the platoon during the 60 minutes}
\label{fig:hway_det_60min}
\end{minipage}
\end{figure}

We now discuss some suggestive graphical depictions for stealthy speed modifications. A formation of 20 vehicles moving at 130 km/h is considered during a time-frame $T=60 min$. The corrupted vehicles are randomly spaced on the lane (see the initial headway $b$ in the plots that follow) accounting for a randomized corruption rate of 3\%. The smoothness factor is set to $\sfact=10$. First, in Figure \ref{fig:traj_det_60min} we show the trajectory of vehicles during 60 minutes. The detail in Figure \ref{fig:traj_det_60min} shows that during the last 5 minutes the vehicles are equally spaced (the distance between them is the target $b$). Figure \ref{fig:speed_60min} shows the speed gain and Figure \ref{fig:speed_per_60min} the speed gain in percents during the 60 minutes, it is only in the 30-th minute that the last vehicle has a speed gain of ~20\% reaching about $154 km/h$, for the rest of the vehicles the speed gain is lower. Figure \ref{fig:hway_60min} shows the evolution of headways between vehicles and Figure \ref{fig:hway_det_60min} gives a detail on this, the headway quickly drops when the speed increases in the middle of the interval. 

Speed modifications as previously depicted appear smooth and may stay stealthy to human drivers. Deciding how stealthy they are requires further studies in the area of human perception and is out of reach for the current work.

\section{Conclusion}

Despite the numerous attacks reported so far, adversarial behaviour has not been previously included in traffic models nor does it appears to be considered in the numerous safety technologies embedded in modern cars. As long as cars are not fully secure, adversarial behaviour is a realistic concern. 
Even small delays in the reaction time due to adversarial actions, e.g. delayed taillights, or small variations in vehicle speed, e.g., by speedometer modifications, can have serious consequences.  We have emphasized this in our models for chain-collisions and provided metrics for adversarial effects by the infinite-collision bound and the instant-reaction-collision speed gain. Proof-of-concept map overlays have shown the effects of such manipulations on more realistic situations.
Finally, our discussion on intelligent adversarial behaviour proves that it is within reach for adversaries to coagulate compromised cars in adversarial platoons that can be further exploited in creating chain collisions. 
Due to the lack of maturity for in-vehicle security technologies, modelling adversarial behaviour for vehicles in traffic should be considered in anticipation of attack scenarios. We hope that our work paves way in this direction.

~~\\

\section*{Acknowledgement} 
This work was supported by the CSEAMAN project a grant of the Romanian National Authority for Scientific Research and Innovation, CNCS-UEFISCDI, project number PN-II-RU-TE-2014-4-1501 (2015-2017). \url{http://www.aut.upt.ro/~bgroza/projects/cseaman}

\bibliographystyle{abbrv}
\bibliography{tmodel}

\begin{IEEEbiography}{Bogdan Groza}
is an associate professor at  Politehnica University of Timisoara (UPT) since 2014. He received his Dipl.Ing. and Ph.D. degrees from UPT in 2004 and 2008 respectively. In 2016 he successfully defended his habilitation thesis having as core subject the design of cryptographic security for vehicular systems. His research interests in embedded systems security and cryptography are reflected by more than 50 publications in conferences or journals  in the field. He regularly serves as member in international conferences committees, reviewer for journals in this area and has directed or participated in several national and international research projects in this field. He was actively involved inside UPT with the development of laboratories by Continental Automotive and Vector Informatik, two world-class manufacturers of automotive software. Currently, he leads the CSEAMAN project, a 2 year research program (2015-2017) in the area of automotive security.
\end{IEEEbiography}

\end{document}